\documentclass[%
pra%
 ,twocolumn%
 ,secnumarabic%
,amssymb, amsmath,nobibnotes, aps, prl]{revtex4-1}

\usepackage{amsmath}
\usepackage[utf8]{inputenc}
\usepackage{amssymb}
\usepackage{amsthm}
\usepackage{dsfont}
\usepackage{graphicx}
\usepackage{dsfont,amsthm,enumerate,amsfonts}
\usepackage{float}
\usepackage[vcentermath]{youngtab}
\usepackage[english]{babel}

\newcommand{\bra}[1]{\ensuremath{\left\langle#1\right|}}
\newcommand{\ket}[1]{\ensuremath{\left|#1\right \rangle}}
\renewcommand{\bra}[1]{\ensuremath{\langle#1|}}
\renewcommand{\ket}[1]{\ensuremath{|#1 \rangle}}

\newcommand{\braket}[2]{\ensuremath{\left\langle#1|#2\right\rangle}}

\newcommand{\parenth}[1]{\ensuremath{({#1})}}
\newcommand{\lamtup}{\boldsymbol{\lambda}}
\newcommand{\rtup}{\boldsymbol{r}}
\newcommand{\xtup}{\boldsymbol{x}}
\newcommand{\nutup}{\boldsymbol{\nu}}

\newcommand{\ptup}{\boldsymbol{p}}
\newcommand{\qtup}{\boldsymbol{q}}
\newcommand{\stup}{\boldsymbol{s}}

\newcommand{\Thetatup}{\boldsymbol{\Theta}}
\newcommand{\omegatup}{\boldsymbol{\omega}}

\newcommand{\ctup}{\boldsymbol{c}}
\newcommand{\ltup}{\boldsymbol{l}}

\begin{document}

\title{Universal and distorsion-free entanglement concentration of multiqubit quantum states in the W class}
\author{Alonso Botero}%
	\affiliation{Departamento de F\'isica, Universidad de los Andes, Cra 1 No 18A-12, Bogot\'a, Colombia}
\author{Jos\'e Mej\'ia}%
	\affiliation{Departamento de F\'isica, Universidad de los Andes, Cra 1 No 18A-12, Bogot\'a, Colombia}


	


\begin{abstract}
We propose a multipartite extension of Matsumoto and Hayashi's  distortion-free entanglement concentration protocol,  which takes  $n$ copies of a general multipartite state and, via  local measurements, produces a  maximally-entangled multipartite  state between local spaces of dimensions $\sim 2^{n E_i}$, where $E_i$ are the local entropies of the input state. However, the extended protocol is generally not universal in the sense that for the same measurement outcomes, the  output state will still depend on the input state. 
Our main result is that when specialized to any state in the multiqubit W class, the protocol is also universal, so that as in the biparatite version, the  output is a unique, maximally-entangled state for each given set of measurement outcomes. Our analysis brings to the forefront a new and interesting family of maximally-entangled multipartite states, which we term Kronecker states. A recurrence relation to obtain the coefficients of the W-class Kronecker states is also given. 
\end{abstract}
\pacs{03.67.Ac,03.65.Ud}

\maketitle

\section{Introduction}
An important problem in quantum information is to determine the extent to which the entanglement of  many copies  of a quantum state shared by several parties can be concentrated with minimal loss into a more compressed, maximally-entangled form, using only local operations and classical communication (LOCC).  In the bipartite case, the problem was essentially solved for pure states in the asymptotic limit by  Bennett et al \cite{bennett_concentrating_1996,bennett_exact_2000}  who proved that  entanglement concentration into copies of a basic entanglement unit, the EPR pair, can be achieved reversibly at an optimal asymptotic rate given by the so-called entanglement entropy.  However, a similar approach to  optimal entanglement concentration in the multipartite setting has proved to be considerably more challenging. As opposed to the bipartite case, multipartite entangled states cannot in general be reversibly transformed into EPR pairs \cite{bennett_exact_2000},  even asymptotically \cite{linden_reversibility_2005}, and the quest for the so-called minimal reversible entanglement generating set (MREGS) \cite{bennett_exact_2000}, which would presumably serve as the  multipartite entanglement units, has so far proved elusive \cite{galvao_tripartite_2000,acin_structure_2003}. Given these difficulties, it may be worth exploring multipartite  concentration schemes where the compressed target states are not  necessarily tensor copies of  fundamental entanglement units.

In the bipartite case, one such scheme is  the \emph{universal distortion-free entanglement concentration} protocol of Matsumoto and Hayashi (MH) \cite{matsumoto_universal_2007,hayashi_error_2003}.  In contrast to standard concentration schemes,  the  protocol extracts from $n$ copies of any bipartite state $\ket{\psi}$, with entanglement entropy $E(\psi)$,  a \emph{single} copy of a bipartite state that is always guaranteed to be maximally-entangled,  of Schmidt rank $\sim2^{n  E(\psi)}$ asymptotically. This makes it attractive to explore a multipartite generalization of the protocol, 
as concentration now refers to the local ranks of a target maximally-entangled state, a notion that is more portable to the multipartite setting than that of the singlet, or more generally, MREG, yield.  In addition, the MH protocol is adapted to a  symmetry that is also present in the multipartite case, namely that of the tensor product $\ket{\psi}^{\otimes n}$ under permutations of the copies. 
The symmetry, which  is made explicit via  Schur-Weyl duality, enters in the protocol through local  projections onto subspaces transforming irreducibly under the symmetric group $S_n$; this  guarantees that the protocol is universal in that no information about the Schmidt basis of the state $\ket{\psi}$ is required, and distortion-free in that the targets are always the maximally-entangled $S_n$-invariant states  residing in tensor products of $S_n$ irreducible modules. 
Such $S_n$-invariant states and the symmetry-adapted local projections  can easily be extended  to a multipartite setting, although whether the resulting protocol remains universal  is a question that would need to be revisited. 

The purpose of  this  paper is  to examine the extension of the MH concentration protocol  to the multipartite setting, with the  the question of universality in mind. Using  Schur-Wey duality and $S_n$-symmetry adapted local measurements, we first show that as in the bipartite case, it is possible to obtain from $n$ copies of an $N$-party state $\ket{\psi}$, a  state residing in the $S_n$-invariant sector of certain tensor products of irreducible $S_n$ modules. Such states, which we term \emph{Kronecker states}, are maximally-entangled in the multipartite sense, as defined by \cite{de_vicente_maximally_2013}, and so in particular  are maximally-entangled when viewed as bipartite states between a single party and the rest. Also, as in the bipartite protocol,  the  rate exponents characterizing  the asymptotic local ranks of these output states are given by the corresponding marginal entropies of $\ket{\psi}$.  However, in contrast to the bipartite case, we also show that  generically, there is a residual indeterminacy in the output state that makes the protocol  non-universal. Our second and main result identifies a class of multiqubit states for which this  residual randomness is absent, namely the class of states that are SLOCC-equivalent to the W-state 
\begin{equation}
\ket{W} \propto \ket{10\dots 0} + \ket{01\dots 0} + \cdots +\ket{00\dots 1}.
\end{equation}
Our main conclusion is therefore that the universality of the bipartite protocol extends to any state in the  W class. 
The proof of our main result is based on a unique simplification that ensues when the set of \emph{SLOCC covariants}, which can in principle be used to separate  SLOCC orbits, is restricted to the W class. For this class, the SLOCC covariants can be computed explicitly.  The explicit knowledge of the W-class covariants makes it possible to efficiently compute the coefficients of the corresponding Kronecker vectors. 

The paper is structured as follows: In section II, we briefly review the multilocal Schur-Weyl decomposition of the n-fold tensor product of multipartite Hilbert spaces. In section III we introduce the Kronecker states as  invariant states in a tensor product of $S_n$ irreducible representations. In section IV we discuss the multipartite extension of the MH protocol, and show that for general SLOCC classes, the protocol is not universal. Theorem 1 in Section V  states our main result: the universality of the extended MH protocol when restricted to the W class. Section VI discusses the machinery of SLOCC covariants and, as shown in Theorem 2, their restriction to the W class;  Theorem 1 then follows as a straightforward consequence of this second theorem. In Section VII we show how to explicitly compute the various states that appear in the Schur-Weyl decomposition of an $n$-fold tensor product of a given W-class state, according to the results of Theorem 1. In particular, we introduce the recurrence relation from which the $W$-class Kronecker state coefficients can be computed efficiently. Some conclusions are given in Section VIII.

\section{Mathematical preliminaries}  We begin by developing the appropriate symmetry-adapted decomposition for tensor products $\ket{\psi}^{\otimes n}$ of   an $N$-partite state $\ket{\psi}\in \mathcal{H}$, where $\mathcal{H} = \bigotimes_{i=1}^{N} \mathcal{H}^{(i)}$ and  $i$ labels the parties; for simplicity we assume that for all $i$,  $\mathcal{H}^{(i)} \cong \mathbb{C}^d$ for some $d \geq 2$.  For a single copy,  reversible local quantum operations are  described by elements $g$ of the \emph{local group} $GL_d^{\times N}$, where $g =\otimes_{i=1}^{N} g^{(i)}$ and the $g^{(i)}$ are elements of  $GL_d$, the linear complex group in $d$ dimensions. Two states $\ket{\psi}, \ket{\phi} \in \mathcal{H}$ are then said to be SLOCC equivalent \cite{bennett_exact_2000} if $\ket{\phi} = g \ket{\psi}$, for some $g \in GL_d^{\times N}$; as usual, local unitary (LU) equivalence refers to equivalence under $g \in U_d^{\times N}$ ($g^{(i)}\in U_d$). For multiple copies, the corresponding space  $\mathcal{H}^{\otimes n}$ can also be viewed as an $N$-partite system, with local spaces $(\mathcal{H}^{(i)})^{\otimes n}$.  The action of  the local group $GL_d^{\times N}$ can then be extended to $\mathcal{H}^{\otimes n}$,  with each $g^{(i)}$  acting as ${(g^{(i)})}^{\otimes n}$ on its corresponding local space.  In addition,  there is a natural action of the permutation group $S_n$  on any given local space,  which on a product basis
is given by $\pi: \ket{e_1e_2\cdots e_n} \to  \ket{\pi^{-1}(e_1e_2\cdots e_n)}$.  By {Schur-Weyl duality} \cite{goodman_symmetry_2009}, $(\mathbb{C}^{d})^{\otimes n} $ decomposes into $GL_d \times S_n$ irreducible representations (irreps) as
\begin{equation}
(\mathbb{C}^{d})^{\otimes n} = \bigoplus_{\lambda\vdash_{\!d} n}V_\lambda\otimes[\lambda],
\end{equation}
where $V_\lambda$ and $[\lambda]$  are  the $GL_d$ and $S_n$ irreps respectively, and where both representations are labeled by integer partitions  $\lambda =(\lambda_1,\lambda_2, \ldots ,\lambda_{d'})$ of $n$ of at most $d$ parts (denoted by $\lambda\vdash_{\!d} n$), with $\sum_{i=1}^r\lambda_i =n$, $\lambda_i \geq \lambda_{i+1} > 0$, and $d'\leq d$ for $GL_d$. 
We note that for large $n$  and fixed $d$,   $\mathrm{dim} V_\lambda$ grows polynomially in $n$ \cite{christandl_spectra_2006},  whereas $\mathrm{dim} [\lambda]$ grows exponentially, with a rate exponent asymptotically approaching the Shannon entropy of the so-called reduced partition $\overline{\lambda} = \lambda/n$. More precisely, we have   
\cite{matsumoto_universal_2007}
\begin{equation}
\label{boundim}
    \left|\frac{1}{n}\log \mathrm{dim}[\lambda] -  H(\overline{\lambda}) \right| = O(\log n/n).
\end{equation}
Now, applying Schur-Weyl duality to each of the local spaces in $\mathcal{H}$, we obtain the decomposition
\begin{equation}
\label{eqWedderburn2}
(\mathcal{H})^{\otimes n} = \bigoplus_{\lamtup} V_{\lamtup}\otimes[\lamtup],
\end{equation} 
where $V_{\lamtup}\equiv  \bigotimes_{i=1}^{N} V_{\lambda^{(i)}}$,    $[\lamtup] \equiv \bigotimes_{i=1}^{N} [\lambda^{(i)}]$, and $\lamtup =  (\lambda^{(1)}, \cdots, \lambda^{(N)})$ with all $\lambda^{(i)} \vdash_d n$,  which achieves a decomposition of the $(\mathcal{H})^{\otimes n}$ into irreps of $GL_d^{\times N} \times S_n^{\times N}$. However,  the tensor product  $\ket{\psi}^{\otimes n}$ is invariant when the \emph{same} permutation is applied to all parties. This means that in fact, 
\begin{equation}\label{psitens}
 \ket{\psi}^{\otimes n} \in \bigoplus_{\lamtup} V_{\lamtup}\otimes[\lamtup]^{S_n},
\end{equation}
where $[\lamtup]^{S_n}$ is the subspace of  $[\lamtup]$  of all  invariant vectors under this coordinated $S_n$ action.    
The dimension of $[\lamtup]^{S_n}$ 
is given by a \emph{generalized Kronecker coefficient}
 \begin{equation}
 k_{\lamtup}=\dfrac{1}{n!}\sum_{\pi \in S_n}\chi_{ \lambda^{(1)}}(\pi) \ldots \chi_{\lambda^{(N)}}(\pi) ,
 \end{equation}
where $\chi_{\lambda}(\pi) $ are the $S_n$ characters. For  $N = 2$,  $k_{\lambda \mu} = \delta_{\mu \nu}$ from $S_n$ character orthogonality, and for $N >3$, $k_{\lamtup}$ can be expanded in terms of the standard ($N=3$) Kronecker coefficients $ k_{\lambda \mu \nu}$ \cite{fulton_representation_2004}, using the $S_n$ character formula $\chi_\lambda(\pi) \chi_\mu(\pi) = \sum_\nu k_{\lambda \mu \nu} \chi_\nu(\pi)  $. For fixed $N$ and $d$, the generalized Kronecker coefficient grows polynomially in $n$. This follows from the asymptotics of the standard Kronecker coefficients \cite{manivel_asymptotics_2015}, and of the number of partitions of $n$ with fixed number of parts \cite{knessl_partition_1990}, both of which are polynomial in $n$.

\section{Kronecker states} We will henceforth refer to any normalized state $ \ket{\mathcal{K}_{\lamtup}} \in [\lamtup]^{S_n}$ as a \emph{Kronecker state}. When considered as entangled states in the $N$-party tensor product space $[\lamtup]$, Kronecker states are the natural distortion-free  target states in  the multipartite generalization of the MH protocol, as follows from the lemma:  

\emph{Lemma 1.--}
For any normalized vector $\ket{\mathcal{K}_{\lamtup}} \in [\lamtup]^{S_n}$, let $\rho_{i}(\mathcal{K}_{\lamtup}) \in \mathcal{L}([\lambda^{(i)}])$ be the one-party  density matrix obtained by  tracing $\ket{\mathcal{K}_{\lamtup}}\bra{\mathcal{K}_{\lamtup}}$ over $[\lambda^{(i)}]^C =\otimes_{j\neq i}[\lambda^{(j)}] $. Then all $\rho_{i}(\mathcal{K}_{\lamtup})$ are multiples of the identity.

The lemma follows from the $S_n$ invariance of Kronecker vectors, which extends to the reduced matrices  $\rho_{i}(\mathcal{K}_{\lamtup})$, together with Schur's lemma. Therefore, all Kronecker states share the unique properties that follow from having maximally-mixed marginals: from the Kempf-Ness theorem \cite{kempf_length_1979,briand_complete_2003}, any two such states  are either LU-equivalent, or else SLOCC-inequivalent; they are  maximally entangled in the multipartite sense of belonging to the maximally entangled set (MES) of states as defined in \cite{de_vicente_maximally_2013} (up to LU equivalence);  clearly,  they are also maximally entangled  with respect to any bipartition involving one party and the rest,
with entanglement entropy scaling with $n$ as $E_{i}(\mathcal{K}_{\lamtup}) \simeq n H(\overline{\lambda})$ asymptotically, as follows from \eqref{boundim}.
  
\section{Generalized MH Protocol}\label{sec:generalizedmhprotocol} The multipartite extension of the MH protocol is based on equation \eqref{psitens}.
 Choosing an orthonormal  basis  $\left\{\,  \ket{\mathcal{K}_{\lamtup,s}}  \,  \right\}$ ($s=1 \cdots k_{\lamtup}$), for each  $ [\lamtup]^{S_n}$, the  general form for  the  expansion of $\ket{\psi}^{\otimes n}$ is then
\begin{equation}\label{eqThreequbits}
\ket{\psi}^{\otimes n}=\bigoplus_{\lamtup: k_{\lamtup}\neq 0 } \left[ 
\sum\limits_{s=1}^{k_{\lamtup}}\ket{\Phi_{\lamtup,s}(\psi)}\otimes\ket{\mathcal{K}_{\lamtup,s}} \right],
\end{equation}
where the $\ket{\Phi_{\lamtup,s}(\psi)}$ are unnormalized states spanning a subspace of $ V_{\lamtup}$ of dimension at most
$ k_{\lamtup}$. As in the bipartite protocol,  
each  party then performs a   measurement  of the set of projectors $\{ P_{\lambda^{(i)}}| \lambda^{(i)} \vdash n\}$   onto the  subspaces $V_{\lambda^{(i)}} \otimes [\lambda^{(i)}]$  of the local product spaces $(\mathcal{H}^{(i)})^{\otimes n}$.  This implements a global measurement of the projectors $ P_{\lamtup} = \bigotimes_{i=1}^N P_{\lambda^{(i)} }$, onto  the subspaces $V_{\lamtup} \otimes [\lamtup]$ in \eqref{eqWedderburn2}. Thus,   $\ket{\psi}^{\otimes n}$ is projected to one of the terms in \eqref{eqThreequbits}: 
\begin{equation}\label{collapse}
\ket{\psi}^{\otimes n} \stackrel{  P_{\lamtup} } {\longrightarrow}
\sum\limits_{s=1}^{k_{\lamtup}}\ket{\Phi_{\lamtup,s}(\psi)}\otimes\ket{\mathcal{K}_{\lamtup,s}}, 
\end{equation}
with probability 
\begin{equation}
  p(\lamtup|\psi) = \left\| P_{\lamtup} \ket{\psi}^{\otimes n} \right\|^2= \sum_{s=1}^{k_{\lamtup}} \braket{\Phi_{\lamtup,s}(\psi)}{\Phi_{\lamtup,s}(\psi)  }.
\end{equation} While this probability will be hard to compute in general, it suffices by the Keyl-Werner theorem \cite{keyl_estimating_2001} that the \emph{marginal} probabilities $p(\lambda^{(i)}|\psi)$  exhibit asymptotic concentration-of-measure around the reduced partition $\overline{\lambda^{(i)}} \simeq r^{(i)}$, where $r^{(i)}$ is  the spectrum of the  partial density matrix $\rho_i$ of $\ket{\psi}$ for the local Hilbert space $\mathcal{H}^{(i)}$, with the eigenvalues  arranged in non-decreasing order. Extending the estimation theorem of  \cite{christandl_spectra_2006} to the $N$-party case,  we then have that for any ball $B_\epsilon(\rtup)  = \{ \boldsymbol{r}' :|r'^{(i)} - r^{(i)}| _1< \epsilon, \forall i\}$ around the local spectra $\rtup =(r^{(1)}, \cdots, r^{(N)})$, there is an $n_0$ such that the reduced partitions satisfy
\begin{equation}
P( \overline{\lamtup} \notin B_\epsilon(\rtup)) < N \epsilon, \qquad \forall n \geq n_0.
\end{equation}
Consequently, the projection \eqref{collapse}  yields $\overline{\lamtup}$ arbitrarily close to $ \rtup$ with unit probability as $n \rightarrow \infty$. Thus, across any bipartition involving one party and the rest, the   per-copy entanglement yields $E_{i}(\mathcal{K}_{\lamtup,s})/n$ of the  states $\ket{\mathcal{K}_{\lamtup,s}}$ resulting from \eqref{collapse} asymptotically tend to  the corresponding bipartite entanglement entropies $E_{i}(\psi)$ of $\ket{\psi}$. 

\begin{figure}
    \centering
    \includegraphics[scale=0.32]{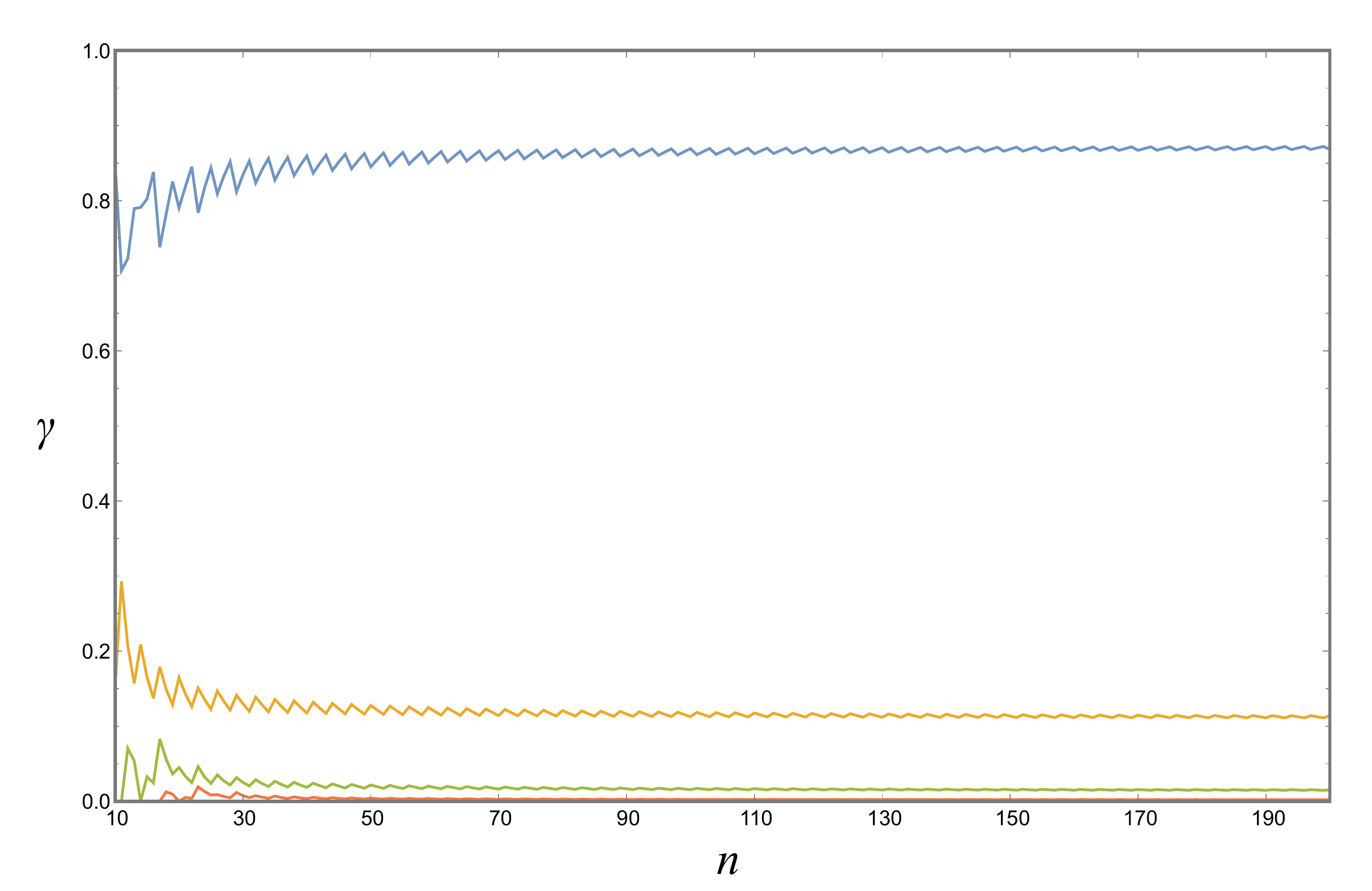}
    \caption{The first four Schmidt coefficients  $\gamma_1 \geq \cdots \geq \gamma_4$ of the state $\sum\limits_{s=1}^{k_{\lamtup}}\ket{\Phi_{\lamtup,s}(\psi)}\otimes\ket{\mathcal{K}_{\lamtup,s}}$ as a function of $n$, for the state $\sqrt{\frac{2}{3}}\ket{000} + \frac{1}{\sqrt{3}}\ket{111} $ and the  partitions $\lambda^{(1)} = \lambda^{(2)}= \lambda^{(3)} = (n- [n/3],[n/3])$, where $[x]$ is the closest integer to $x$. The graph suggests that the largest Schmidt coefficient $\gamma_1$ converges to a numerical value lower than  1, and hence that the protocol is not even  approximately universal in the asymptotic limit.}
    \label{fig:kron}
\end{figure}

There is a caveat,  however.  The universality of the bipartite MH protocol rests on the fact that the resulting state from \eqref{collapse} is a \emph{separable} state  $\ket{\Phi_{\lamtup}}\ket{\mathcal{K}_{\lamtup}}$, where  the target  $\ket{\mathcal{K}_{\lamtup}}$ is the maximally entangled state
between the spaces $[\lambda^{(1)}] = [\lambda^{(2)}]$. Thus, the target  state is readily obtained by simply discarding the $V_{\lamtup}$ space, in which the state $\ket{\Phi_{\lamtup}}$ has $ O(\log n)$ entanglement. From the viewpoint of the multipartite protocol, this is due to the fact that the bipartite Kronecker coefficient is $k_{\lamtup} \leq 1$. 
But more  generally, $k_{\lamtup} >1$ for $N>2$,  so  the projection \eqref{collapse}  will generally yield a state with a residual entanglement  between  $V_{\lamtup}$ and $[\lamtup]^{S_n}$ with Schmidt rank of at most $k_{\lamtup}$, and therefore  $O(\log n)$ entanglement entropy.
Figure \ref{fig:kron} illustrates this fact for  the tripartite  GHZ class; indeed, we have verified numerically that the residual entanglement  has the maximal Schmidt rank  $k_{\lamtup}$ for this class of states  (see Appendix A for  details on the techniques used to obtain these Schmidt coefficients).  
A pure Kronecker vector can therefore only be obtained  by   performing an additional  set of  local measurements on the individual $V_{\lambda^{(i)}}$ spaces in order to break the entanglement between $V_{\lamtup}$ and $[\lamtup]^{S_n}$. For  each set of outcomes of these additional measurements,  the resulting state will be a linear superposition of the $\ket{\mathcal{K}_{\lamtup,s}} $, with coefficients that will generally depend on $\ket{\psi}$. This means that in general, the protocol is not  universal, since we can only produce  Kronecker states randomly from an ensemble that depends on $\ket{\psi}$ and the outcomes of the additional measurement. Moreover, as Kronecker states are generically not locally interconvertible, it will generally  be impossible to obtain, by local means, a unique target Kronecker state for each set of outcomes in the total measurement  sequence.

\section{Universality in the W-class}
 Interestingly, it turns out  that for states in  certain  non-trivial SLOCC classes, the Schmidt rank of  the projected state \eqref{collapse} can be one, even if $k_{\lamtup}>1$. Our main result is that  this is  the case for states 
 in the $N$-qubit W class:

\emph{Theorem 1.--}Let $\ket{\psi}$ be a state in the multipartite W SLOCC-class, so that 
$\ket{\psi} = g \ket{W}$ for some $g \in GL_2^{\times N}$. Then, the
 multilocal Wedderburn decomposition of $\ket{\psi}^{\otimes n}$ simplifies to the form  \begin{equation}
 \label{Wmulti}
\ket{\psi}^{\otimes n}=\bigoplus_{\lamtup \in \Lambda^{(W)}_n}
\ket{{\Phi}_{\lamtup}(\psi)}\otimes\ket{\mathcal{K}^{(W)}_{\lamtup}} 
\end{equation}
where $\Lambda^{(W)}_n$ is the  set of  $\lamtup$  with all $\lambda^{(i)} \vdash_{2}\! n$ for which the reduced second rows 
$\overline{\lambda}_2^{(i)} =  \lambda^{(i)}_2/n$ satisfy  
\begin{equation}\label{admiss}
 2\overline{\lambda}_2^{(i)} \leq  \sum_{i=j}^{N} \overline{\lambda}_2^{(j)} \leq 1, 
 \end{equation}
and  each $\ket{\mathcal{K}^{(W)}_{\lamtup}} $ is a unique Kronecker vector in  $[\lamtup]^{S_n}$ that is common to the whole W class. 

\begin{figure}
    \centering
     \includegraphics[scale=0.35]{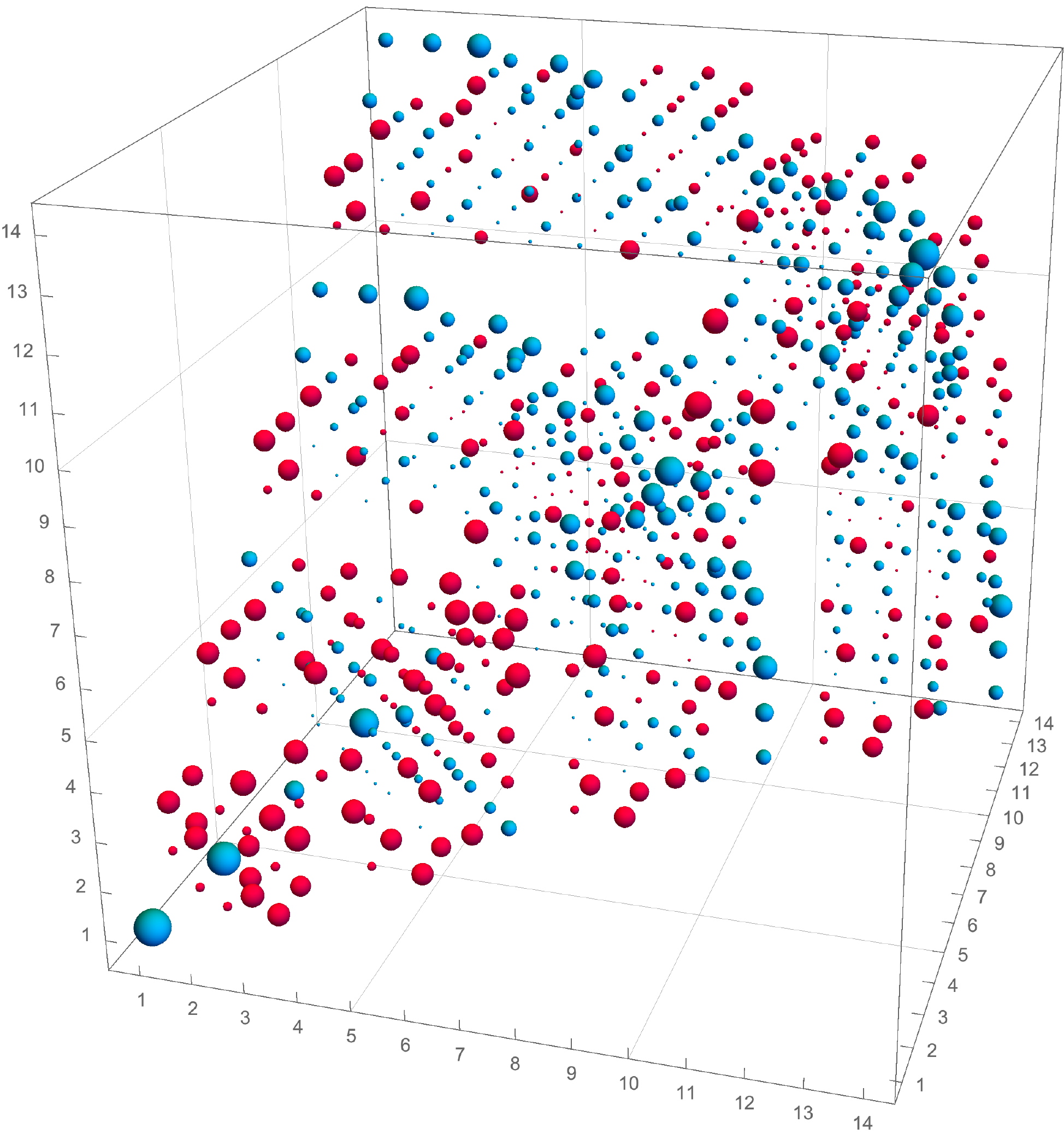}
    \caption{Graphical representation of the $N=3$  Kronecker state $\ket{\mathcal{K}^{(W)}_{\lamtup}} $ for $n=7$ and a triplet for which $k_{\lamtup}=2$: $\lambda^{(1)}=\lambda^{(2)}=\lambda^{(3)} = (5,2) $, with $\mathrm{dim}[(5,2)]=14$. The labels correspond to the  elements of the each $[\lambda^{(i)}]$ basis, ordered lexicographically. Each sphere  represents  the coefficient for the corresponding product basis element, with the radius representing the magnitude and the color representing the sign.  }
    \label{fig:wkron}
\end{figure}

Thus,  Matsumoto and Hayashi's universal distortion-free protocol extends \emph{mutatis mutandi} to all multiqubit states in the W class, with a  unique target maximally entangled multipartite state $\ket{\mathcal{K}^{(W)}_{\lamtup}} $ obtained in the projection  \eqref{collapse} (Fig. \ref{fig:wkron} provides a graphical representation of one such state). Note that the extension encompasses all entangled states in the case $N=2$, which are  SLOCC-equivalent to the two-qubit W state. The conditions in  \eqref{admiss}  ensure that the support of $p(\lamtup|\psi)$ is compatible with the  correspondence between partitions and marginal spectra, since  replacing  $\overline{\lambda}^{(i)}$ by the  spectra $ r^{(i)}$, the leftmost inequality in  \eqref{admiss} gives the marginal spectral condition satisfied by all  $N$-qubit states \cite{higuchi_one-qubit_2003}, while the rightmost inequality is the  generalization of an additional spectral condition satisfied by W class states \cite{walter_entanglement_2013}.

Theorem 1 is  a corollary of a second theorem we present concerning the restriction to the W class of the ring of so-called multiqubit \emph{SLOCC covariants} \cite{briand_complete_2003}, which are closely related to  the possible states $\ket{\Phi_{\lamtup,s}(\psi)}$ that can appear in the decomposition \eqref{eqThreequbits}.

\section{ W class SLOCC covariants} \label{sec:wclasscovariants}

To establish the connection wth the SLOCC covariants, let  $\ket{\psi}$ and $\ket{\psi'}$ be unnormalized states such that $ \ket{\psi'} = g \ket{\psi} $ for some $g \in SL_2^{\times N}$ (all $g^{(i)}$ with unit determinant). Then, 
the  $\ket{\Phi_{\lamtup,s}(\psi)}$ in \eqref{eqThreequbits} satisfy
\begin{equation}\label{transphi}
\ket{\Phi_{\lamtup}(\psi)}=D_{\lamtup}(g^{-1})\ket{\Phi_{\lamtup}(\psi')},
\end{equation}
where $D_{\lamtup}$ is the $GL_{2}^{\times N}$ representation matrix for $V_{\lamtup}$ and we  omit the  index $s$. Now,  a $GL_2$ irrep $V_{\lambda}$ when restricted to  $SL_2$,  is isomorphic to the space of homogeneous polynomials $P_{\nu}(x)$ of degree $\nu=\lambda_1 -\lambda_2$ in  indeterminates $x \equiv (x_0, x_1)^T$, with coefficients   transforming equivalently to components of $V_{\lambda}$ vectors   under the  action $ (g , f(x) ) \rightarrow f( g^Tx)$ \cite{fulton_representation_2004}; specifically,  with the correspondence 
\begin{equation}\label{mappoly}
 \ket{\nu,\omega}\  \leftrightarrow\  m_{\nu,\omega}(x) = \sqrt{{ {\nu \choose \omega} }}  x_0^{\nu-\omega} x_1^\omega ,
\end{equation}
  coefficients in the  basis  $m_{\nu,\omega}(x) $ transform with the same  matrix as those of $V_\lambda$ vectors   with respect to an  $SL_2$ highest-weight basis $\ket{\nu,\omega}$ (the standard  angular momentum basis $\ket{j,m}$ with $j = \nu/2$ and $m = \nu/2 - \omega$).  Thus we may associate
to any  $\ket{\Phi_{\lamtup}(\psi)}$ a so-called SLOCC-\emph{covariant} $I_{\Phi}(\psi,\xtup)$-- a multihomogeneous polynomial  in $\psi$ and  $N$   auxiliary variables $x^{(i)} =(x^{(i)}_0,x^{(i)}_1)^{T}$, satisfying
\begin{equation}
I_{\Phi}(\psi',\xtup') = I_{\Phi}(\psi,\xtup),
\end{equation} 
where    $(g^{(i)})^T{x^{(i)} }'= x^{(i)}$ \cite{briand_complete_2003}. The covariant is  of  multidegree $(n, \nutup)$,  where $n$ is the degree in $\psi$ and $\nutup$ is the tuple  of  degrees in the auxiliary variables $\xtup$, with $\nu_i = \lambda_1^{(i)} - \lambda_2^{(i)}$.

Now, it is known that the ring of SLOCC covariants is finitely-generated, and that a generating set can be obtained in principle using Cayley's Omega process \cite{olver_classical_1999}, also know as the process of \emph{iterated transvectants}, adapted to the multiqubit case \cite{briand_complete_2003}. The process starts from the \emph{base form} associated with the state,
\begin{equation}
A_{\psi}(\xtup)=\sum_{\boldsymbol{\imath} \in \{0,1\}^N}\psi_{i_1\cdots i_N}x^{(1)}_{i_1}\cdots x^{(N)}_{i_N},
\end{equation}
and iteratively generates new covariants from old ones   through their transvectant, defined as
\begin{equation}\label{transvect}
(F,G)^{\ltup}=\left.\Omega^{l_1}_{x^{(1)}}\cdots\Omega^{l_N}_{x^{(N)}}F(\xtup)G(\xtup')\right|_{\xtup'=\xtup},
\end{equation}
where the $\Omega$ operator is
\begin{equation}
\Omega_x=\dfrac{\partial}{\partial x_0} \dfrac{\partial}{\partial x_1'}- \dfrac{\partial}{\partial x_1} \dfrac{\partial}{\partial x_0'}.
\end{equation}

We will show that from the base form  corresponding to any state in the W class, the  process generates at most one linearly independent covariant for a multidegree $(n, \nutup)$. To this end, we use the fact that  any  state in the W class is, up to LU transformations,  completely specified  by its marginal spectra \cite{sawicki_when_2013}, and 
is LU-equivalent to the state \cite{yu_multipartite_2013} 
\begin{equation}\label{ecstatepsi}
\ket{\psi} = \sqrt{c^{(0)}}|\boldsymbol{0} \rangle + \sum_{i=1}^{N}\sqrt{c^{(i)}}|\boldsymbol{1}_i \rangle, 
\end{equation}
where $\boldsymbol{0}$ is a sequence of all zeros and $\boldsymbol{1}_i$  is a sequence with a ``1" at position $i$ and otherwise all zeros,  and  $c^{(i)}$ are real with $\sum_{i=0}^{N}c^{(i)}=1$. Thus, for states in the class, it suffices to use the base form
\begin{equation}\label{baseform}
A_{\psi}(\xtup) =   \left[\sqrt{c^{(0)}}+ \sum_{i=1}^{N}\sqrt{c^{(i)}}\frac{x_1^{(i)}}{x_0^{(i)}} \right]x_0^{(1)}...x_0^{(N)}.
\end{equation}
With the  notation  $\boldsymbol{a}^{\boldsymbol b}  \equiv \prod_{i=1}^{N}\left(a^{(i)}\right)^{b^{(i)}}$, we then have:

\emph{Theorem 2.--} Any non-vanishing covariant of multidegree $(n, \nutup)$, generated from the base form $A_{\psi}$ through the process of iterated transvectants must be such that: i) $(\nu^{(i)} - n) \mod 2 =0$ ii) $w \geq 0$, and iii) $\nu^{(i)} \geq w$, where $w= \frac{1}{2}\left[\sum_{i=1}^{N}\nu^{(i)} - (N-2)n\right]$, in which case the covariant is numerically proportional to
\begin{equation}\label{wcov}
I_{\psi}^{(n, \nutup)} (\xtup) = \ctup^{(n\boldsymbol{e}-\nutup)/4}  \xtup_0^{ \boldsymbol{\nu} -w \boldsymbol{e} } A_{\psi}(\xtup)^{w},
\end{equation}
where   $\boldsymbol{e}\equiv (1,1,\ldots, 1)$.

To prove the theorem we  use  the fact that  if $F(\xtup)$ and $G(\xtup)$ are multihomogeneous  of multidegrees $\boldsymbol{f}$ and $\boldsymbol{g}$ in the auxiliary variables, then their transvectant \eqref{transvect} will also be multihomogeneous functions of multidegree $\boldsymbol{f} + \boldsymbol{g} - 2\ltup$. Following Olver \cite{olver_classical_1999}, we can then adapt the transvectant  to functions of the projective coordinates $p_i = x_1^{(i)}/x_0^{(i)}$ associated with  multihomogeneous functions. Explicitly,  for any $F(\xtup)$  of multidegree $\boldsymbol{f}$,  its \emph{projective form}  $\widehat{F}(\ptup)$ is defined by \begin{equation}
F(\xtup) =  \boldsymbol{x}_0^{\boldsymbol{f}}\, \widehat{F}(x_1^{(i)}/x_0^{(i)},\ldots, x_1^{(N)}/x_0^{(N)}),   
\end{equation}
so in particular, the base form \eqref{baseform} has projective form 
$
\widehat{A_{\psi}}(\ptup) = \sqrt{c^{(0)}} + \sum_{i=1}^{N} \sqrt{c^{(i)}} p_i.
$
For simplicity, consider a transvectant involving a single pair of auxiliary variables; then following \cite{olver_classical_1999}, the projective  transvectant
is given by
\begin{equation}
    \widehat{(F,G)^{l_i}}=  \sum_{k=0}^{l_{i}}c(l_i,k,f_i,g_i) \frac{\partial^{l_i-k}\widehat{F}}{\partial p_i^{l_i-k}} \frac{\partial^{k}\widehat{G}}{\partial p_i^{k}},
\end{equation}
where $c(l,k,f,g) =l! (-1)^k {f- l +k \choose k}{g- k  \choose l-k}$, provided $0 \leq l_i \leq \min(f_i,g_i)$, and otherwise vanishes.  Now 
suppose that $\widehat{F} =  a  \widehat{A_{\psi}}^t$ and $\widehat{G} = b \widehat{A_{\psi}}^s$, with $t,s \geq 0$ and  $a$ and $b$  independent of the $p_i$. Then,
 $
 \widehat{(F,G)^{l_i}} \propto a b (c^{(i)})^{l_i} \widehat{A_{\psi}}^{t + s - l_i},
$
where the proportionality constant is numerical and may be equal to zero. The multivariable generalization of this result is straightforward:  
\begin{equation}\label{projtrans}
    \widehat{(F,G)^{\ltup}}(\ptup) \propto  a b \ctup^{\ltup} \widehat{A_{\psi}}^{t + s - \sum_i l_i} \, .
 \end{equation}
Hence, the projective forms of all  covariants generated from the base form  \eqref{baseform} through the Omega process  are always proportional to a power of $\widehat{A_{\psi}}$. Re-expressing the covariants in their homogeneous forms, we  find that any  covariant derived from the base form $A_{\psi}(\xtup)$ must then be of the form
\begin{equation}
    I \propto  \ctup^{\ltup/2} \boldsymbol{x}_0^{l\boldsymbol{e} - 2\rtup} A_{\psi}^{m-l}
\end{equation}
for some $\ltup$ and $m \geq l$, where $l= \sum_i l_i$. This can be checked inductively by noting that \eqref{baseform} is of this form, with $(\ltup = \boldsymbol{0}$, $m=1)$, and that the transvectant  preserves the form. Equation \eqref{wcov} in theorem 2 is obtained by setting $m=n$,  $\nutup = n\boldsymbol{e} - 2 \ltup$, and $w=n-l$ to match the degrees.  Solving for the $l_i$ we obtain $l_i = (n- \nu_i)/2$ and $w$ as defined in  theorem. Finally, condition \emph{i)} follows from  the fact that the $l_i$ are integers, and conditions \emph{ii)} and \emph{iii)} from the fact that iterated transvectants that start with $A_{\psi}$ cannot generate negative powers in the $x_0^{(i)}$ or the $x_1^{(i)}$. 

Theorem 1 then follows from the correspondence between the states $\ket{{\Phi}_{\lamtup} (\psi)} \in V_{\lamtup}$ and covariants of degrees $(n,\nutup)$. The  product form \eqref{Wmulti} is a consequence of there being at most  one linearly independent covariant  for $\nu_i = \lambda_1^{(i)} - \lambda_2^{(i)}$ for states of the W class, which determines, up to a constant and LU-equivalence, the state $\ket{{\Phi}_{\lamtup} (\psi)}$.  The  admissible set $\Lambda_W(n)$ in \eqref{admiss} follows from conditions \emph{ii)} and \emph{iii)} in Theorem 2. Note that the parity condition  \emph{i)} is automatically satisfied. 

\section{The states $\ket{\Phi_{\lamtup}(\psi)}$ and $\ket{\mathcal{K}^{(W)}_{\lamtup}}$ of the W class}\label{computations}

Having  established the decomposition \eqref{Wmulti} for multiqubit  W-class states, in this section we address the question of the  explicit form of the states $\ket{\Phi_{\lamtup}(\psi)}$ and the target Kronecker states $\ket{\mathcal{K}^{(W)}_{\lamtup}}$. We shall work in the Schur-Weyl basis that arises naturally from the so-called Schur transform \cite{bacon_efficient_2006}, which we briefly describe in the next subsection. The coefficients of the  state  $\ket{\Phi_{\lamtup}(\psi)}$ are readily obtained from the results of Theorem 2 up to normalization (Eq. \eqref{phigen}). This explicit form can then be used to obtain a recurrence relation for the coefficients of $\ket{\mathcal{K}^{(W)}_{\lamtup}}$ using the recurrence relations of the Schur transform (Eqs. \eqref{recrelk} and \eqref{fexp}). 

\subsection{The Schur-Weyl Basis}\label{sec:schurweylbasis}
Following \cite{bacon_efficient_2006}, the basis adapted to the single-party $d=2$ Schur-Weyl decomposition 
$
(\mathbb{C}^2)^{\otimes n} = \bigoplus_{\lambda \vdash_2 n}V_\lambda \otimes [\lambda],
$
 will be denoted by $\ket{\lambda,\omega,q} = \ket{\lambda,\omega} \otimes \ket{\lambda, q}$, where $\ket{\lambda,\omega}$ and $\ket{\lambda,q}$ are basis elements for the subspaces $V_{\lambda}$ and $[\lambda]$ respectively. The  $\ket{\lambda,\omega, q }$ are elements of a standard angular momentum basis $\ket{j,m; q}$ with degeneracy index $q$, where  $j = \frac{1}{2}(\lambda_1 - \lambda_2)$, \ $m =  \frac{n}{2}- \omega $ with $\lambda_1 \geq \omega \geq \lambda_2$, and  $q$   labels the different copies of the corresponding spin-$j$ representation  in $(\mathbb{C}^2)^{\otimes n}$. Let   $\ket{s} $ denote  the  elements of the standard computational basis, labeled by binary sequences $  s = (s_1 s_2 \cdots s_n) \in \{0,1\}^n $ corresponding to sequences of ``up or down spins''; then   $\ket{\lambda,\omega,q}$ is a linear combination of states $\ket{s}$ with isotypical   sequences of Hamming weight $\omega$.   In turn,  the label $q$ represents a sequence of partitions
$
\lambda(1)  \rightarrow  \lambda(2) \rightarrow \cdots \rightarrow \lambda(n-1)     \rightarrow \lambda(n)
$
 that is traversed as the spins in the computational basis are successively added, using   standard angular-momentum addition. Here $\lambda(1)=(1,0)$, $\lambda(n)=\lambda$, and $\lambda(k)$ is a two-part partition of $k$  that is obtained   by adding a one to either  the first or, if allowed, to the second part of $\lambda(k-1)$ (in the language of Young diagrams, by adding a box to either the first or second row). We  therefore take $q=(q_1q_2 \cdots q_n)$, with $q_1=0$, to be a binary sequence encoding  allowed partition sequences, where  $q_k$ takes the value 0 (resp. 1) if $\lambda(k)$ was obtained by adding a one to the first (resp. second) part of $\lambda(k-1)$. Intermediate partitions are therefore given by $\lambda(k) = (\sum_{j=1}^{k}(1-q_j),\sum_{j=1}^{k}q_j)$. 

 Using this procedure, the Schur-Weyl basis can be constructed recursively. To simplify notation, let $(\lambda', \omega', q')$
 be labels for Schur-Weyl for $\mathbb{C}^{n-1}$, such that $q'$ is obtained from the sequence $q$  by omitting the last element $q_n$, $\omega'$ is either $ \omega$ or $\omega+1$, and    $\lambda' = (\lambda_1 -(1-q_n), \lambda_2 -q_n)$. 
 The $\mathbb{C}^n$ Schur-Weyl basis element $\ket{\lambda,\omega,q}$ is then obtained using angular momentum addition of a $\mathbb{C}^{n-1}$ state $\ket{\lambda',\omega',q'}$ and a single qubit at the $n$th register according to
 \begin{equation}
 \label{recstate1}
     \ket{\lambda,\omega,q} = \sum_{s_n \in \{0,1\}} \Gamma^{\lambda ,\omega}_{q_n, s_n} \ket{\lambda',\omega'=\omega-s_n,q'} \otimes \ket{s_n },
 \end{equation}
 where  $\Gamma^{\lambda ,\omega}_{ q_n,s_n}$ are matrix elements (with row/column indices $\in \{0,1\}$)
of the matrix
\begin{equation}
\label{CGmat} 
    \Gamma^{\lambda ,\omega} = \left( \begin{array}{cc} \sqrt{ \frac{\lambda_1-\omega}{\lambda_1 - \lambda_2}} & \sqrt{ \frac{\omega-\lambda_2}{\lambda_1 - \lambda_2}} \\ {} & {} \\
    \sqrt{ \frac{\omega-\lambda_2+1}{\lambda_1 - \lambda_2+2}} & -\sqrt{ \frac{\lambda_1-\omega+1}{\lambda_1 - \lambda_2+2}} \end{array} \right )
\end{equation}
of  Clebsch-Gordan coefficients $\braket{j_1,j_2;m_1,m_2}{j m}$ for 
$ j = \frac{1}{2}(\lambda_1 - \lambda_2)$, $ m = \frac{n}{2} - \omega$, $j_1 = j - (-1)^{q_n}/2$, $m_1 = m -(-1)^{s_n}/2$, $j_2 = 1/2$, and $m_2 = (-1)^{s_n}/2$, with elements on the first row defined  to be zero if $\lambda_1 = \lambda_2$.  
Appealing to the reality of Clebsch-Gordan coefficients, we let $B^{\lambda ,\omega,q}_s = \braket{s}{\lambda, \omega, q} = \braket{\lambda, \omega, q}{s}$ be the transformation matrix elements between the computational and Schur-Weyl bases for $\mathbb{C}^n$;  then,  \eqref{recstate1} entails the recurrence relation for the coefficients $B^{\lambda ,\omega,q}_s$ for $\mathbb{C}^n$ and those for $\mathbb{C}^{n-1}$:
\begin{equation}
\label{reccoefs1}
 B^{\lambda ,\omega ,q }_{s } = \Gamma^{\lambda ,\omega}_{q_n, s_n} B^{\lambda' ,\omega' ,q'}_{s'}, \qquad B^{(1,0) ,s_1 ,(0)}_{(s_1)} =1,
 \end{equation}
where  $s'$  is obtained from $s$ by omitting the last element $s_n$ and $\omega'= \omega-s_n$. Note that $B^{\lambda ,\omega,q}_s$ is zero unless $\lambda= (\sum_{j=1}^{n}(1-q_j),\sum_{j=1}^{n}q_j)$ and $\omega = \sum_{j=1}^{n} s_j$.

For the $N$-partite Schur-Weyl basis for $((\mathbb{C}^2)^{\otimes n})^{\otimes N} = \bigoplus_{\lamtup }V_{\lamtup} \otimes [\lamtup] $, we use the notation $\ket{\lamtup,\omegatup,\qtup} =\ket{\lamtup,\omegatup} \otimes \ket{\lamtup, \qtup}$, where as before,  boldface symbols denote  $N$-tuples (e.g., $\qtup = (q^{(1)} \cdots q^{(N)}) $, etc.) of the corresponding single party variables and $\ket{\lamtup,\omegatup} = \otimes_{i=1}^{ N}  \ket{\lambda^{(i)},\omega^{(i)}}$, etc.  Similarly,  defining 
\begin{equation}
        B^{\lamtup,\omegatup,\qtup}_{\stup}=\prod_{i=1}^{N}B^{\lambda^{(i)},\omega^{(i)},q^{(i)}}_{s^{(i)}}, \qquad  \Gamma^{\lamtup,\omegatup}_{ \qtup_n,  \stup_n },=\prod_{i=1}^{N}\Gamma^{\lambda^{(i)} ,\omega^{(i)}}_{ q_n^{(i)}, s_n^{(i)}},
\end{equation}
 the $ B^{\lamtup,\omegatup,\qtup}_{\stup}$ are the expansion coefficients in the computational basis $\{ \ket{\stup } \}$ of $((\mathbb{C}^2)^{\otimes n})^{\otimes N}$, with $\ket{\stup } = \ket{s^{(1)}}\ket{ s^{(2)} } \ldots \ket{s^{(N)}}$. The recurrence relation \eqref{reccoefs1} then generalizes to
 \begin{equation}
\label{reccoefs2}
 B^{\lamtup,\omegatup,\qtup}_{\stup} = \Gamma^{\lamtup,\omegatup}_{ \qtup_n,  \stup_n } B^{\lamtup',\omegatup',\qtup'}_{\stup'},
 \end{equation}
 where the previous relation between primed and unprimed labels applies to each party separately.

\subsection{ The states $\ket{\Phi_{\lamtup}(\psi)} $}
As discussed earlier, any state in the W class is  LU-equivalent to a state $\ket{\psi}$ of the form \eqref{ecstatepsi}, where the coefficients $c^{(i)}$ can be regarded as implicit functions of the marginal spectra of the state. Hence, up to LU-equivalence, the state $\ket{\Phi_{\lamtup}(\psi)}$ is proportional to the state in correspondence with the covariant $I_\psi^{(n, \nutup)}$ in \eqref{wcov} using the mapping \eqref{mappoly} between covariants and $SL_2$ basis elements. The mapping can also be expressed in terms of the $GL_2$ basis elements in the Schur-Weyl basis,  by noting that the $SL_2$ weights are obtained by subtracting $\lambda_2$   from the $GL_2$  weights, so that:
\begin{equation}\label{mappoly2}
 \ket{\lambda,\omega}\  \leftrightarrow\  \sqrt{\frac{(\lambda_1 - \lambda_2)!}{(\lambda_1 - \omega)!(\omega- \lambda_2)!} }  x_0^{\lambda_1-\omega} x_1^{\omega-\lambda_2} .
\end{equation}
Since $\ket{\Phi_{\lamtup}(\psi)}$ can only be determined from the covariant  $I_\psi^{(n, \nutup)}$ up to a normalization, it will be convenient   to define a fiducial, also unnormalized state $\ket{\widehat{\Phi}_{\lamtup}(\psi)}$ through  the correspondence 
\begin{equation}
\ket{\widehat{\Phi}_{\lamtup}(\psi)} \leftrightarrow  \frac{\sqrt{ \prod_{i=1}^N (\lambda_1^{(i)}-\lambda_2^{(i)})!}}{w!} I_{\psi}^{(n,\boldsymbol{\nu})}(\boldsymbol{x}), 
\end{equation}
where $I_{\psi}^{(n,\boldsymbol{\nu})}(\boldsymbol{x})$ is as defined in \eqref{wcov}. 
Using the definition \eqref{baseform} of the base form $A_{\psi}(\xtup)$, expanding as a polynomial in the auxiliary variables, and recalling that  $\nu^{(i)} = \lambda_1^{(i)} - \lambda_2^{(i)}$  and $w= n - \sum_i \lambda_2^{(i)}$, we obtain the expansion
\begin{equation}
\label{phigen}
    \ket{\widehat{\Phi}_{\lamtup}(\psi)}=\\
   \sum_{\omega^{(0)}=0}^{n} \dfrac{(c^{(0)})^{\omega^{(0)}/2} } {\omega^{(0)}!} \sum_{\omegatup} {\ctup}^{\omegatup/2} \sqrt{A_{\lamtup,\omegatup}}
 \ket{ \lamtup, \omegatup},
\end{equation}
where  the sum is over all  weights $\omega^{(i)}$ such that $\sum_{i=0}^{N} \omega^{(i)} = n$ and $\lambda_1^{(i)} \geq \omega^{(i)} \geq \lambda_2^{(i)}$ for $1 \leq i \leq N$, and
\begin{equation}
\label{defas}
   A_{\lamtup,\omegatup}=\prod_{i=1}^{N} A_{\lambda^{(i)},\omega^{(i)}}, \qquad   A_{\lambda,\omega}= \dfrac{(\lambda_1- \omega)! }{( \omega-\lambda_2)! } .
\end{equation}
Note that for the W state, $c^{(0)}=0$, $c^{(i)} = 1/\sqrt{N}$, and hence
\begin{equation}
\label{phiw}
    \ket{\widehat{\Phi}_{\lamtup}(W)}=
    \frac{1}{N^{n/2}}\sum_{\omegatup}  \sqrt{A_{\lamtup,\omegatup}}
 \ket{ \lamtup, \omegatup},
\end{equation}
where the sum is over  weights such that $\sum_{i=1}^{N} \omega^{(i)} = n$.

To determine the state $\ket{\Phi_{\lamtup}(\psi)} $, it suffices to find the the proportionality constant $\eta_{\lamtup}$ such that
$
    \ket{\Phi_{\lamtup}(\psi)} =  \eta_{\lamtup} \ket{\widehat{\Phi}_{\lamtup}(\psi)},
$
which is independent of the state $\ket{\psi}$; therefore,  the constant can be expressed in  terms of quantities involving the  $W$-state, namely
\begin{equation}
\label{defeta}
\eta_{\lamtup} =\sqrt{ \frac{p(\lamtup|\psi)}{Z_{\lamtup}(\psi)}}= \sqrt{ \frac{p(\lamtup|W)}{Z_{\lamtup}(W)}}, 
\end{equation}
where $Z_{\lamtup}(\psi )= \| \ket{\widehat{\Phi}_{\lamtup}(\psi)} \|^2$ and $p(\lamtup|\psi) = \| \ket{{\Phi}_{\lamtup}(\psi)} \|^2 $. Thus, since $Z_{\lamtup}(\psi )$ and $Z_{\lamtup}(W )$ can be computed from \eqref{phigen} and/or \eqref{phiw}, the constant $\eta_{\lamtup}$ and hence  the   probabiltities  $p(\lamtup|\psi) = \| \ket{{\Phi}_{\lamtup}(\psi)} \|^2 $ can be obtained for a general state once the corresponding probability
 $p(\lamtup|W)$ for the $W$ state is known. This probability can in principle be obtained from the results the next subsection.

\subsection{The states $\ket{{\mathcal{K}}^{(W)}_{\lamtup}}$ }

With the explicit knowledge of the states $\ket{\widehat{\Phi}_{\lamtup}(\psi)}$, it is then possible to obtain a recurrence relation for the coefficients of a state proportional to $\ket{{\mathcal{K}}^{(W)}_{\lamtup}}$ in the $[\lamtup ]$ basis. For this, let us cast the expansion   \eqref{Wmulti} for the $W$ state as
\begin{equation}
    \ket{W}^{\otimes n}=\bigoplus_{\lamtup \in \Lambda^{(W)}_n}
\ket{\widehat{{\Phi}}_{\lamtup}}\otimes\ket{\widehat{\mathcal{K}}_{\lamtup}} 
\end{equation}
where $\ket{\widehat{{\Phi}}_{\lamtup}} =\ket{\widehat{\Phi}_{\lamtup}(W)}$ as defined in \eqref{phiw} and  $\ket{\widehat{\mathcal{K}}_{\lamtup}} = \eta_{\lamtup}\ket{{\mathcal{K}}^{(W)}_{\lamtup}} $, with $\eta_{\lamtup}$  as defined in \eqref{defeta}. Then, the Schur Weyl decomposition of $ \ket{W}^{\otimes n}$ can be written in terms of that  of $\ket{W}^{\otimes n-1}$ as
\begin{equation}
  \bigoplus_{\lamtup \in \Lambda^{(W)}_n}\!
\ket{\widehat{{\Phi}}_{\lamtup}}\ket{\widehat{\mathcal{K}}_{\lamtup}} = \left(\bigoplus_{\lamtup' \in \Lambda^{(W)}_{n-1}}
\! \ket{\widehat{{\Phi}}_{\lamtup'}}\ket{\widehat{\mathcal{K}}_{\lamtup'}}\right) \otimes \ket{W}. 
\end{equation}
Letting $\ket{\widehat{\mathcal{K}}_{\lamtup}} = \sum_{\qtup}\widehat{\mathcal{K}}_{\lamtup,\qtup}\ket{\lamtup,\qtup}$, using expansion \eqref{phiw} and the recurrence relation \eqref{reccoefs2}, we obtain the recurrence relation between the coefficients of $\ket{\widehat{\mathcal{K}}_{\lamtup}}$ for $n$ in terms of those for $n-1$:
\begin{equation}
\label{recrelk}
  \widehat{\mathcal{K}}_{\lamtup,\qtup} = F_{\lamtup,\qtup} \widehat{\mathcal{K}}_{\lamtup',\qtup'},
\end{equation}
where primed and unprimed quantities are related as in \eqref{reccoefs2} and
\begin{equation}
\label{fdef}
 F_{\lamtup,\qtup}=\sum_{\stup_n \in B_W} \sqrt{ \frac{{A}_{\lamtup',\omegatup'}}{{A }_{\lamtup,\omegatup}} } \Gamma^{\lamtup ,\omegatup}_{\qtup_n, \stup_n}  ,
\end{equation}
where $B_W = \{\boldsymbol{1}_1, \cdots \boldsymbol{1}_N\}$ is the set of binary sequences in the W state. Note that this equation must be independent of the  weights $\omegatup$  if the weights satisfy the condition $\sum_{i=1}^{N} \omega^{(i)} = n$ of  expansion \eqref{phiw}. 
Using  \eqref{CGmat} and \eqref{defas}, we can show that for any given party,
\begin{equation}
\sqrt{\frac{ A_{\lambda' , \omega'}}{ A_{\lambda, \omega}}}{\Gamma}^{\lambda,\omega}_{q_n,s_n}  = \frac{ 1 + s_n[\omega - q_n(\lambda_1 -1) -(1-q_n)\lambda_2 ]}{\sqrt{\lambda_1 -\lambda_2 + 2 q_n}} \, , 
\end{equation}
where $\lambda'= \lambda-(1-q_n,q_n)$, and $\omega'= \omega - s_n$. We therefore see that the numerator differs from $1$ only  when $s_n=1$. Replacing into \eqref{fdef} and using the facts that $\stup_n$ runs over all sequences where only one of the entries has $s_n=1$ and  $\sum_{i=1}^N \omega^{(i)} = n$, we finally obtain
the proportionality $ F_{\lamtup,\qtup}$ constant in the recurrence relation  \eqref{recrelk}:
\begin{equation}
\label{fexp}
 F_{\lamtup,\qtup}=\frac{ n- \sum_i^{N}q_n^{(i)}  (\lambda_1^{(i)} + 1)-\sum_i^{N} (1-q_n^{(i)})\lambda_2^{(i)}}{\sqrt{\prod_{i=1}^N( \lambda_1^{(i)} - \lambda_2^{(i)} + 2 q_n^{(i)})}}\, ,
\end{equation}
where it is understood that the coefficient vanishes whenever the denominator vanishes. As expected, this coefficient is independent of the weights.

Once $\ket{\widehat{\mathcal{K}}_{\lamtup}}$ is obtained from the recurrence relation $\eqref{recrelk}$,  we have
$
\eta_\lambda=  \| \ket{\widehat{\mathcal{K}}_{\lamtup}} \| ,
$
and the states $\ket{\Phi_{\lamtup}(\psi)} $ and $\ket{{\mathcal{K}}_{\lamtup}}$ are then completely determined, as are the probabilities $p(\lamtup|\psi)$. An alternative method to compute these probabilities exactly is presented in Appendix B. 

Figure \ref{fig:wkron} illustrates the set of coefficients obtained  using \eqref{recrelk} and \eqref{fexp} for $N=3$ and $n=7$. Some explicit coefficient values for $N=3$ and $N=4$ and $n \leq 5$ are also  given in the supplementary material.

\section{Conclusion} 
In summary, we have shown that  the multipartite extension of the HM protocol is able to produce maximally entangled multipartite states with exponentially large local ranks described by asymptotic rates  given by the  von Neummann entropies of the reduced one-party density matrices of the state. We have also shown that while  the  multipartite protocol is generally not universal, it remains universal within the class of multiqubit W states. In proving our result, we have obtained the explicit form of all non-vanishing SLOCC covariants for multiqubit states in the W class, which for a given multidgree are unique up to a constant. Our result identifies in the   
 Kronceker states $\ket{\mathcal{K}^{(W)}_{\lamtup}}$  a new family of large-rank, maximally entangled multipartite states, the coefficients of which can be recursively computed with a simple  algorithm. The interesting entanglement and combinatorial properties of these states  may prove useful for quantum information tasks.

Our main result establishes  the universality of  the  multipartite   MH protocol when restricted to the W class, and provides a way of computing all elements involved in the Schur-Weyl decomposition \eqref{Wmulti},  including the probability $p(\lamtup|\psi) = \braket{\Phi_{\lamtup}(\psi)}{\Phi_{\lamtup}(\psi)}$.  Additionally,  we provide in  Appendix B, an  alternative  formula to compute the probability $p(\lamtup|W)$, which can then be used to obtain  $p(\lamtup|\psi)$ for a general W-class state using relation \eqref{defeta}. However, none of these results are practical to further characterize the asymptotic concentration of measure of  $p(\lamtup|\psi)$  beyond what can be inferred from the Keyl-Werner theorem. It therefore remains an open question as to what is the explicit form of the 
 rate function $R({\overline{\mathbb{\lamtup}}} |\psi) = \lim_{n \rightarrow \infty} \frac{1}{n} \log 
p(\lamtup|\psi) $
 that exactly characterizes this  concentration of measure  in the same way that the relative entropy $D(\overline{\lambda}|r_\psi)$ does in the bipartite case.


Looking further,   our results suggest  an intriguing connection between the SLOCC class of a general state $\ket{\psi}$ and the residual entanglement of the states $\sum\limits_{s=1}^{k_{\lamtup}}\ket{\Phi_{\lamtup,s}(\psi)}\otimes\ket{\mathcal{K}_{\lamtup,s}}$   arising in the  MH multipartite protocol. We believe that a better understanding of this connection  may shed an additional light on the nonlocal properties of different SLOCC classes and their relation to the general problem of asymptotic interconvertibility of multipartite entangled states. 

\acknowledgements

AB would like to thank M. Christandl for helpful discussions. 

\appendix 
\section{Gram matrix and Schmidt coefficients}
In this appendix we  show how to explicitly compute the Schmidt coefficients of the state $\sum\limits_{s=1}^{k_{\lamtup}}\ket{\Phi_{\lamtup,s}(\psi)}\otimes\ket{\mathcal{K}_{\lamtup,s}}$ for $N$-qubit GHZ class states of the form 
\begin{equation}
    \ket{\psi_{GHZ}}=\sqrt{1-\alpha}\ket{00\cdots0}+\sqrt{\alpha}\ket{11\cdots1}
\end{equation}
 with $0<\alpha<1$. These computations can be used to show that for GHZ states the Schmidt rank is indeed larger than one as discussed in section \ref{sec:generalizedmhprotocol}  and that the Schmidt coefficients, when arranged in decreasing value, appear to show an exponential decay that is independent of $n$ as shown in Figure \ref{fig:kron}. 
 \\
 
 We begin with the $n$-th tensor product of $\ket{\psi_{GHZ}}$, which can be expanded in the product basis of the $N$ parties in terms of identical sequences $s$ as
\begin{equation}\label{ecghz}
    \ket{\psi_{GHZ}}^{\otimes n}=\sum_{\omega}\sum_{s\sim \omega}\xi^{\omega}(\alpha)\ket{s}\cdots\ket{s},
\end{equation}
where $\omega$ is the Hamming weight of each sequence and $\xi^{\omega}(\alpha)=\alpha^{\frac{\omega}{2}}(1-\alpha)^{\frac{n-\omega}{2}}$. Now,  performing a multilocal Schur transform,  the expansion of the state in the  multilocal Schur-Weyl basis becomes, as discussed in section \ref{sec:schurweylbasis},
\begin{equation}\label{ecghz2}
    \ket{\psi_{GHZ}}^{\otimes n}=\sum_{\lamtup,\omega,\qtup}
\xi^{\omega}(\alpha)\sum_{s\sim \omega}B^{\lamtup,\omegatup,\qtup}_{s}\ket{\lamtup,\omega^{\times N},\qtup},
\end{equation}
where $\ket{\lamtup,\omega^{\times N},\qtup}=\bigotimes_{i=1}^{N}\ket{\lambda^{(i)},\omega,q^{(i)}}$ and we use the notation $s \sim \omega$ to denote sequences $s$ with Hamming weight $\omega$. 
This can be written in a manner  similar to \eqref{eqThreequbits}, namely, 
\begin{equation}
    \ket{\psi_{GHZ}}^{\otimes n}=\sum_{\lamtup}\sum_{\omega}\xi^{\omega}(\alpha)\ket{\lamtup,\omega^{\times N}}\ket{K^{\lamtup}_{\omega}},
\end{equation}
where $\ket{\lamtup,\omega^{\times N}}$ are the basis vectors of $V_{\lamtup}$ with the same weight $\omega$ in each party, and 
\begin{equation}
    \ket{K^{\lamtup}_{\omega}}=\sum_{\qtup}\sum_{s\sim \omega}B^{\lamtup,\omega^{\times N},\qtup}_{s}\ket{\lamtup,\qtup} 
\end{equation}
are the unnormalized Kronecker states $\in [\lamtup]^{S_n}$ relative to each $\ket{\lamtup,\omegatup}$, where the weight $\omega$ runs over all values in the range between $max(\lambda_2^{(1)},\ldots,\lambda_2^{(N)}) $ and $min(\lambda_1^{(1)},\ldots,\lambda_1^{(N)}) $.
For a given set of $\lamtup$, the reduced density matrix $\rho$ in the $V_{\lamtup}$ subspace is then
\begin{equation}
    \rho_{V_{\lamtup}}=\sum_{\omega,\omega'}G^{\lamtup}_{\omega\omega'}\ket{\lamtup,\omega^{\times N}}\bra{\lamtup,\omega'^{\times N}},
\end{equation}
where $G^{\lamtup}$ is the Gram matrix with components
\begin{equation}
G^{\lamtup}_{\omega,\omega'}=\dfrac{\xi^{\omega}(\alpha)\xi^{\omega'}(\alpha)\braket{K^{\lamtup}_{\omega}}{K^{\lamtup}_{\omega'}}}{\sum_{\omega}\xi^{\omega}(\alpha)^2\braket{K^{\lamtup}_{\omega}}{K^{\lamtup}_{\omega}}},
\end{equation}
with 
\begin{equation}\label{ecgram}
    \braket{K^{\lamtup}_{\omega}}{K^{\lamtup}_{\omega'}}=\sum_{\qtup}\sum_{s\sim\omega}\sum_{s'\sim\omega'}B^{\lamtup,\omega^{\times N},\qtup}_{s}B^{\lamtup,\omega'^{\times N},\qtup}_{s'}.
\end{equation}
The Schmidt coefficients of $\sum\limits_{s=1}^{k_{\lamtup}}\ket{\Phi_{\lamtup,s}(\psi)}\otimes\ket{\mathcal{K}_{\lamtup,s}}$ are then the eigenvalues $\gamma_i$ of the Gram matrix $G^{\lamtup}$.

The overlaps  $\braket{K^{\lamtup}_{\omega}}{K^{\lamtup}_{\omega'}}$ in the Gram matrix can be computed relatively efficiently as we now show. First, we write \eqref{ecgram} as
\begin{multline}
\label{kinner}
     \braket{K^{\lamtup}_{\omega}}{K^{\lamtup}_{\omega'}}=f_{\lamtup}
     \sum_{s\sim\omega}\sum_{s'\sim\omega'}C^{\lambda^{(1)}}_{\omega,\omega'}({s,s'})\cdots C^{\lambda^{(N)}}_{\omega,\omega'}({s,s'}),
\end{multline}
where  $f_{\lamtup} =\prod_{i=1}^{N}f_{\lambda^{(i)}}$, $f_\lambda=\dim[\lambda]$, and 
\begin{equation}\label{deflouck}
         C^{\lambda}_{\omega,\omega'}(s, s')=\dfrac{1}{f_\lambda}\sum_{q}B^{\lambda,\omega,q}_{s}B^{\lambda,\omega',q}_{s'}.
\end{equation}
Under permutations, the matrix elements $B^{\lambda,\omega,q}_{s}$ transform as
\begin{equation} 
B^{\lambda,\omega,q}_{\pi s} = S^{\lambda}_{q,q'}(\pi) B^{\lambda,\omega,q'}_{s},
\end{equation}
where $S_{q,q'}^{\lambda}(\pi)$ is the representation matrix for $\pi$ in the irrep. $[\lambda]$. Using Schur's grand orthogonality theorem, we can then show that 
\begin{equation}
    C^{\lambda}_{\omega,\omega'}(\pi s,\pi s') = C^{\lambda}_{\omega,\omega'}(s, s').
\end{equation}
 Hence, $C^{\lambda}_{\omega',\omega}(s, s')$ only depends on the type of the joint sequence $(s,s')^{T}=((s_1,s_1')(s_2,s_2')\cdots(s_n,s_n'))$, which can be represented by a $2 \times 2$ \emph{joint sequence weight} matrix $\Theta$, where the matrix elements $\Theta_{ij}$, with $(i,j) \in \{0,1\}^2$, indicate the number of times that the pair $(i,j)$ appears in the joint sequence $(s,s')^{T}$. Therefore, replacing the sum over sequences $s,s'$ with  a sum over all possible joint sequence weights $\Theta$, \eqref{kinner} can be expressed as
\begin{multline}\label{eclouckmulti}
         \braket{K^{\lamtup}_{\omega}}{K^{\lamtup}_{\omega'}}=f_{\lamtup}
         \sum_{\Theta}^{*}\dfrac{n!}{\prod_{ij}\Theta_{ij}!}C^{\lambda^{(1)}}_{\omega,\omega'}(\Theta)\cdots C^{\lambda^{(N)}}_{\omega,\omega'}(\Theta),
     \end{multline}
 where the the asterisk indicates that the sum is restricted to joint sequence weights satisfying the  conditions
\begin{eqnarray}
\Theta_{10} + \Theta_{11} & = & \omega\label{constraints1} \\
\Theta_{01} + \Theta_{11} & = & \omega' \\
\sum_{i,j} \Theta_{ij}  & = & n\label{constraints3}.
\end{eqnarray} 
Note that the term $\frac{n!}{\prod_{ij}\Theta_{ij}!}$ is  the number of  sequence pairs $(s,s')$ with joint weight $\Theta$. Up to a factor of $n!$ the quantities $C^{\lambda^{(i)}}_{\omega,\omega'}(\Theta)$ are the so-called Louck polynomials \cite{chen_combinatorics_1998,louck_unitary_2008}, which are the matrix-valued coefficients in the expansion of the $GL_2$ representation matrix $D^{\lambda}(X)$ in terms of monomials of components of $X\in GL_2$; explicitly in terms of our definition of $C^{\lambda}_{\omega,\omega'}$,
\begin{equation}\label{eqDrep}
         D^{\lambda}_{\omega,\omega'}(X)=\sum_{\Theta}^{*}\dfrac{n!}{\prod_{ij}\Theta_{ij}!}C^{\lambda}_{\omega,\omega'}(\Theta)\prod_{i,j}X_{ij}^{\Theta_{ij}}.
\end{equation}
From the orthogonality and completeness relations of the Schur-Weyl basis, we can also obtain the orthogonality condition
\begin{equation}
    \sum_{\Theta}^{*}\dfrac{n!}{\prod_{ij}\Theta_{ij}!}C^{\lambda}_{\omega\omega''}(\Theta)C^{\lambda'}_{\omega'\omega'''}(\Theta)=\dfrac{1}{f_{\lambda}}\delta_{\lambda\lambda'}\delta_{\omega \omega'}\delta_{\omega''\omega'''},
\end{equation}
and the completeness condition
\begin{equation}
    \sum_{\lambda,\omega,\omega'}f_{\lambda}C^{\lambda}_{\omega\omega'}(\Theta)C^{\lambda'}_{\omega\omega'}(\Theta')=\dfrac{\prod_{ij}\Theta_{ij}!}{n!}\delta_{\Theta\Theta'},
\end{equation}
where in both cases, $\Theta$ is understood to be compatible with the weights $\omega, \omega'$.

From the constraints \eqref{constraints1}-\eqref{constraints3}, the matrix $\Theta$ has only one independent parameter which we choose to be $\Theta_{01}$ and henceforth denote as $x$. The Louck polynomials can then be expressed in terms of the so-called Hahn-Eberlein polynomials \cite{mceliece_new_1977}, which are easily programmable on a computer and are defined as 
\begin{equation}\label{echahn}
    E^{\lambda}_{\omega,\omega'}(x)={}_{3}F_{2}\left(\left.\begin{array}{ccc}
         -\lambda_2,&-x,&j-n-1  \\
         -\omega',&\omega-n 
    \end{array}\right|1\right).
\end{equation}
 The relation between the Louck and the Hahn-Eberlein polynomials reads
\begin{equation}\label{eclouckhahn}
    C^{\lambda}_{\omega,\omega'}(\Theta)=\dfrac{(\omega_{<})!(n-\omega_>)!}{n!}
    \sqrt{\dfrac{A_{\lambda,\omega_<}}{A_{\lambda,\omega_>}}}
    E^{\lambda}_{\omega_{<},\omega_{>}}(x),
\end{equation}
where $\omega_{>}$ (resp. $\omega_{<}$) is the greater (resp. lesser) of $\omega$ and $\omega'$ and $A_{\lambda,\omega}$ is as defined in Eq. \eqref{defas}. Therefore, for fixed weights $\omega, \omega'$,  the sum in \eqref{eclouckmulti} can be taken over $x$, where the constraints on $x$ are such that all matrix elements of $\Theta$ are non-negative and
\begin{equation}
    \dfrac{n!}{\prod_{i,j}\Theta_{ij}}=\dfrac{n!}{x!(n-\omega'-x)!(\omega'-x)!(\omega-\omega'+x)!}.
\end{equation}
\begin{figure}[H]
    \centering
    \includegraphics[scale=0.5]{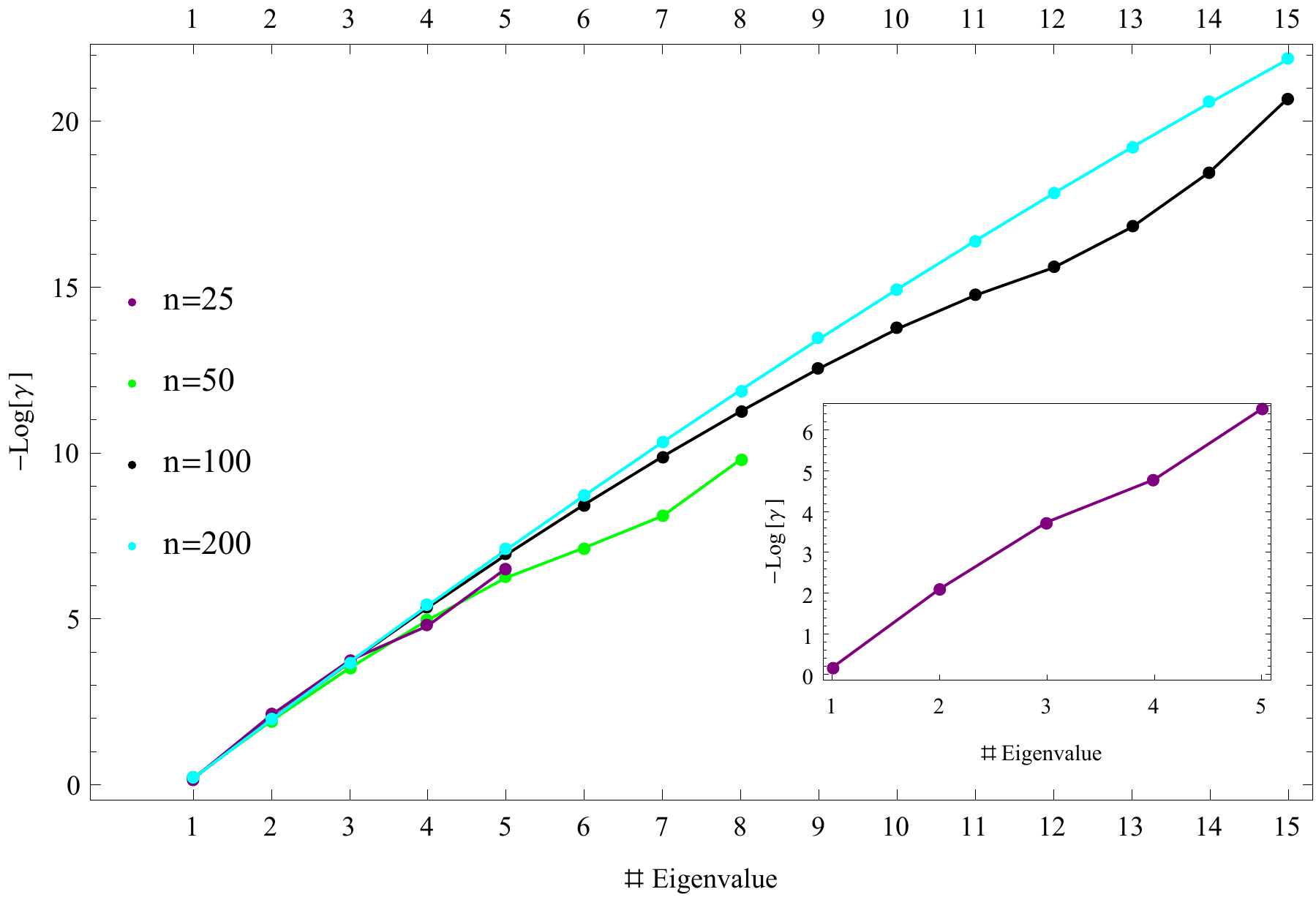}
    \caption{Eigenvalues $\gamma$ of the Gram matrix ordered in decreasing order for different values of $n$. The exponential decay feature is evident from the linearity of the graphs. }
    \label{fig:grameigen}
\end{figure}
The result shown in Figure 1 in the main body of the paper  corresponds to the ranked eigenvalues of the Gram matrix $G^{\lamtup}$ for the case $\alpha=1/3$, and  partitions that are typical according to the Keyl-Werner theorem, so that  the reduced partitions satisfy $\overline{\lambda} \simeq (2/3,1/3)$. Another view of this result is provided by Fig. \ref{fig:grameigen}, which suggests that the residual Schmidt coefficients  exhibit an exponential decay law that appears insensitive to the value of $n$. 

\section{The probability $p(\lamtup|W)$}
In this appendix we will give an expression to explicitly calculate the probability $p(\lamtup|W)$, that together with equation \eqref{defeta}  allows us to calculate $p(\lamtup|\psi)$ for any $\psi$ in the W class. 

Given the $\ket{W}$ state 
\begin{equation}\label{ec:W}
    \ket{W}=\dfrac{1}{\sqrt{N}}\sum_{i=1}^{N}\ket{\boldsymbol{1}_i},
\end{equation}
 (using the notation of section \ref{sec:wclasscovariants}), 
and expanding in the computational basis, we obtain
\begin{equation}\label{ec:W2}
       \ket{W}^{\otimes n}=\dfrac{1}{N^{n/2}}\sum_{\omegatup}\sum_{\stup\sim \omegatup}^{*}\ket{s^{(1)}}\otimes\cdots\otimes\ket{s^{(N)}},
       \end{equation}
where $\omegatup=(\omega^{(1)},\ldots,\omega^{(N)})$ is the tuple of Hamming weights, $\stup=(s^{(1)},\ldots,s^{(N)})$ the tuple of sequences (where $s^{(i)} \sim \omega^{(i)}$), and  the $*$ in the sum represents the constraint  that the sequences $\stup$ must be generated from the $n$-fold tensor product of the W state, i.e.,  $(s^{(1)}_{i}s^{(2)}_i\cdots s^{(N)}_i)\in\{(100\cdots0),(010\cdots0),\ldots,(000\cdots 1)\}$ for all $i$. Performing a multilocal Schur transform, \eqref{ec:W2} becomes
\begin{equation}
    \ket{W}^{\otimes n}=\dfrac{1}{N^{n/2}}\sum_{\omegatup}\sum_{\stup\sim\omegatup}^{*}B^{\lamtup,\omegatup,\qtup}_{\stup}\ket{\lamtup,\omegatup,\qtup},
\end{equation}
so that the probability  $p(\lamtup|W)$ is
\begin{equation}\label{ecproblarga}
         p(\lamtup|W)=\dfrac{1}{N^{n}}\sum_{\omegatup,\qtup}\sum_{\stup\sim\omegatup}^{*}\sum_{\stup'\sim\omegatup}^{*}B^{\lamtup,\omegatup,\qtup}_{\stup}B^{\lamtup,\omegatup,\qtup}_{\stup'}.
     \end{equation}
     Using \eqref{deflouck}, the probability is expressed in terms of  Louck polynomials as
\begin{equation}
    p(\lamtup|W)=\dfrac{1}{N^{n}}f_{\lamtup}\sum_{\omegatup}\sum_{\stup\sim\omegatup}^{*}\sum_{\stup'\sim\omegatup}^{*} C^{\lamtup}_{\omegatup,\omegatup'}(\Thetatup) ,
\end{equation}
with
\begin{equation}
    C^{\lamtup}_{\omegatup,\omegatup'}(\Thetatup)=\prod_{i=1}^{N}C^{\lambda^{(i)}}_{\omega^{(i)},{\omega^{(i)}}'}(\Theta^{(i)}),
\end{equation}
where the  $\Theta^{(i)}$ are  joint sequence weights of the sequences $s^{(i)}$ and $s^{(i)}{}'$. Replacing the sums over the isotypical sequences $\stup,\stup'$ compatible with the state W to a sum over $\Thetatup = (\Theta^{(1)} \ldots \Theta^{(N)})$ we have
   \begin{equation}\label{explicitprobabilityw}
         p(\lamtup|W)=\dfrac{1}{N^{n}}f_{\lamtup}\sum_{\omegatup}\sum_{\Thetatup}Z(\Thetatup,\omegatup)C^{\lamtup}_{\omegatup,\omegatup'}(\Thetatup) ,
     \end{equation}
     where  $Z(\Thetatup,\omegatup)$ is a combinatorial factor counting the number of joint sequences $(s^{(1)},\cdots s^{(N)}, s^{(1)}{}' \cdots s^{(N)}{}')$ such that: 1) every pair $s^{(i)}$ and $s^{(i)}{}'$ is of weight $\omega^{(i)}$ and  of compatible joint weight $\Theta^{(i)}$, and 2) the joint sequence $(s^{(1)},\cdots s^{(N)})$ and $(s^{(1)}{}',\cdots s^{(N)}{}')$ are generated from the W state. These constrains can be written in terms of a tensor $Q$ defined by the $4N$ equations
 \begin{equation}\label{ectensor}
 \Theta^{(i)}_{0,0}=Q_{i,i},\quad \Theta^{(i)}_{1,1}=\sum_{a,b\neq i}Q_{a,b},
 \end{equation}
 \begin{equation}\label{ectensor2}
\Theta^{(i)}_{0,1}=\sum_{a\neq i}Q_{i,a},\quad \Theta^{(i)}_{1,0}=\sum_{a\neq i}Q_{a,i}.
 \end{equation}
Thus $Z(\Thetatup,\omegatup)$ can  be expressed through the $N^2$ components of  $Q$ as
\begin{equation}\label{ec:Z}
    Z(\Thetatup,\omegatup)=n!\sum_{Q_{\text{indep}}}{\prod\limits_{i,j=1}^{N}(Q_{i,j})!^{-1}},
\end{equation}
where the sum runs over the $N^2-3N+1$ independent components of $Q$ compatible with equations \eqref{ectensor} and \eqref{ectensor2}.
Using the independent parameter $x^{(i)}$  for each $\Theta^{(i)}$,  it can be shown that \eqref{ec:Z} can be expressed as the following constant term identity
\begin{equation}\label{ecz}
Z(\Thetatup,\omegatup)=n!\left. \prod_{i=1}^{N}\frac{\left(\frac{\sum_{k\neq i}z_k}{z_i}\right)^{x^{(i)}}}{(\omega^{(i)}-x^{(i)})!x^{(i)}!}\right|_{\text{C.T.}},
\end{equation} where C.T. stands for the term that is constant in all the $z_i$. Using equations \eqref{echahn} and \eqref{eclouckhahn},  to calculate efficiently the Louck polynomials in terms of Hahn-Eberlein polynomials, and \eqref{ecz} to calculate $Z(\Thetatup,\omegatup)$, the probability $p(\lamtup|W)$ can then be explicitly computed using \eqref{explicitprobabilityw}.



\section{Tables of Kronecker state coefficients}

In this supplementary material we present some examples of  W-class Kronecker vector coefficients, obtained using the results of section VII.3 of the main article. For each subspace $[\lambda^{(i)}]$ we  label the basis elements of the corresponding Schur transform  basis $\ket{\lambda^{(i)},q}$ (see Sec. {})  by the ordinal index of the binary sequence $q$ when the set of admissible binary sequences for the partition $\lambda^{(i)} $ is ordered lexicographically. For instance, in Table I  the coefficients of the Kronecker state corresponding to the partition $\lambda^{(i)}=(2,1)$ (for $i=1,2,3$) of $n=3$ copies of the three party W state. In this case the possible binary sequences $q$ are $001$ and $010$ with labels are $1$ and $2$ respectively. Thus, for example, the multipartite label $(1,2,1)$ denotes the coefficient of the term $\ket{\lambda^{(1)},001}\ket{\lambda^{(2)},010}\ket{\lambda^{(3)},001}$.

We also use the standard convention of Clebsch-Gordan tables in which  a square root common to all the coefficients is omitted, with the understanding that a negative sign  appears outside of  the  square root. For the cases presented in Tables \ref{tab:n3-1}-\ref{tab:n5-2}, the Kronecker coefficient $k_{\lamtup}$ is one, so the Kronceker vectors $\ket{\mathcal{K}_{\lamtup}^{(W)}}$  in those cases are common to all SLOCC classes. Tables 
 \ref{tab:n6-1} and \ref{tab:N4-1} present the first non-trivial cases for $N=3$ ($k_{\lamtup}=2$) and $N=4$ ($k_{\lamtup}=4$). In both of these cases, all the $\lambda^{(i)}$ are equal, so the states have an additional permutation symmetry with respect to the parties. For this reason, tables \ref{tab:n6-1} and \ref{tab:N4-1}  omit the terms involving permutations of the parties, which are understood to have the same coefficient.  This brings down the number of terms to be displayed from 192 to 46 in Table \ref{tab:n6-1} and from 29 to 5 in table \ref{tab:N4-1}. 

\onecolumngrid

\begin{table}[H]
    \centering
    \begin{tabular}{|c|c|c|c|}
    \hline
         $\parenth{1,1,1}$&$\parenth{1,2,1}$ & $\parenth{2,1,1}$ &$\parenth{1,1,2}$\\[5pt]   
         \hline
         $\frac{1}{4}$ &$-\frac{1}{4}$ &$-\frac{1}{4}$ &$-\frac{1}{4}$ \\[5pt]
         \hline
    \end{tabular}
    \caption{$N=3$, $n=3$, $\lambda^{(1)}=\lambda^{(2)}=\lambda^{(3)}=(2,1)$}
    \label{tab:n3-1}
\end{table}

\begin{table}[H]
\centering
\begin{tabular}{|c|c|c|c|c|c|c|c|c|c|c|}
\hline
     $\parenth{1,1,1}$& $\parenth{1,2,2} $ & $\parenth{1,3,3}$ & $\parenth{2,1,2}$ & $\parenth{2,2,1}$ & $\parenth{2,2,2}$ & $\parenth{2,3,3}$& $\parenth{3,1,3}$ & $\parenth{3,2,3}$  & $\parenth{3,3,1}$ & $\parenth{3,3,2}$\\[5pt]
     \hline
     $\frac{2}{9}$& $-\frac{1}{18}$& $-\frac{1}{18}$ & $-\frac{1}{18}$&$-\frac{1}{18}$& $\frac{1}{9}$ &$-\frac{1}{9}$&$-\frac{1}{18}$& $-\frac{1}{9}$& $-\frac{1}{18}$& $-\frac{1}{9}$\\[5pt]
     \hline
\end{tabular}
\caption{$N=3$,$n=4$, $\lambda^{(1)}=\lambda^{(2)}=\lambda^{(3)}=(3,1)$}
\label{tab:n4-1}
\end{table}

\begin{table}[H]
\centering
\begin{tabular}{|c|c|c|c|c|c|c|c|}
\hline
     $\parenth{1,2,1}$& $\parenth{1,3,2} $&$\parenth{2,1,1}$&$\parenth{2,2,1}$ & $\parenth{2,3,2}$ & $\parenth{3,1,2}$ & $\parenth{3,2,2}$ &$\parenth{3,3,3}$\\[5pt]
     \hline
     $\frac{1}{6}$ & $\frac{1}{6}$& $\frac{1}{6}$ & $\frac{1}{12}$&$-\frac{1}{12}$& $\frac{1}{6}$  &$-\frac{1}{12}$&$-\frac{1}{12}$\\[5pt]
     \hline
\end{tabular}
\caption{$N=3$,$n=4$, $\lambda^{(1)}=\lambda^{(2)}=(3,1),  \lambda^{(3)}=(2,2)$}
\label{tab:n4-2}
\end{table}

\begin{table}[H]
\centering
\begin{tabular}{|c|c|c|c|c|c|c|c|c|c|c|c|c|}
\hline
     $\parenth{1,1,2}$& $\parenth{1,2,1} $& $\parenth{1,2,2}$ &$\parenth{1,3,3}$&$\parenth{1,4,4}$ & $\parenth{2,1,3}$ &$\parenth{2,2,3}$&$\parenth{2,3,1}$& $\parenth{2,3,2}$&$\parenth{2,3,3}$&$\parenth{2,4,4}$&$\parenth{3,2,3}$ & $\parenth{3,3,2}$\\
     \hline
     $\frac{1}{12}$& $\frac{1}{12}$ & $\frac{1}{45}$& $-\frac{1}{180}$& $-\frac{1}{180}$& $\frac{1}{12}$ & $-\frac{1}{180}$& $\frac{1}{12}$& $-\frac{1}{180}$& $\frac{1}{90}$& $-\frac{1}{90}$ & $\frac{1}{15}$& $\frac{1}{15}$\\[5pt]
     \hline
\end{tabular}

\begin{tabular}{|c|c|c|c|c|c|c|c|c|c|c|c|}
\hline
    $\parenth{3,3,3}$&$\parenth{3,4,4}$ & $\parenth{4,1,4}$ & $\parenth{4,2,4}$& $\parenth{4,3,4}$ & $\parenth{4,4,1}$& $\parenth{4,4,2}$&$\parenth{4,4,3}$ & $\parenth{5,2,4}$ & $\parenth{5,3,4}$ & $\parenth{5,4,2}$& $\parenth{5,4,3}$\\
     \hline
      $\frac{1}{30}$& $-\frac{1}{30}$& $\frac{1}{12}$& $-\frac{1}{180}$  & $-\frac{1}{90}$& $\frac{1}{12}$& $-\frac{1}{180}$& $-\frac{1}{90}$& $\frac{1}{15}$& $-\frac{1}{30}$& $\frac{1}{15}$& $-\frac{1}{30}$\\[5pt]
      \hline
\end{tabular}
\caption{$N=3$,$n=5$, $\lambda^{(1)}=(3,2)$, $ \lambda^{(2)}=(4,1)$, $\lambda^{(3)}=(4,1)$}
\label{tab:n5-1}
\end{table}

\begin{table}[H]
\centering
\begin{tabular}{|c|c|c|c|c|c|c|c|c|c|c|c|c|c|c|c|}
\hline
     $\parenth{1,1,1}$&$\parenth{1,1,2}$ &$\parenth{1,2,3}$&$\parenth{1,3,3}$ &$\parenth{1,4,4}$&$\parenth{1,5,4}$&$\parenth{2,1,3}$ &$\parenth{2,2,1}$& $\parenth{2,2,2}$&$\parenth{2,2,3}$&$\parenth{2,3,2}$&$\parenth{2,3,3}$& $\parenth{2,4,4}$& $\parenth{2,5,4}$& $\parenth{3,3,3}$& $\parenth{3,5,2}$\\[5pt]
     \hline
     $\frac{1}{30}$& $\frac{1}{18}$& $-\frac{1}{72}$& $-\frac{1}{24}$& $-\frac{1}{72}$& $-\frac{1}{24}$ & $-\frac{1}{72}$& $\frac{1}{30}$& $-\frac{1}{72}$ & $\frac{1}{36}$& $-\frac{1}{24}$& $-\frac{1}{48}$ & $-\frac{1}{36}$& $\frac{1}{48}$& $-\frac{1}{24}$ &$-\frac{1}{24}$\\[5pt]
     \hline
\end{tabular}
\begin{tabular}{|c|c|c|c|c|c|c|c|c|c|c|c|c|c|c|c|}
     \hline
     $\parenth{3,2,3}$& $\parenth{3,3,1}$ & $\parenth{3,4,4}$& $\parenth{4,1,4}$&$\parenth{4,2,4}$& $\parenth{4,3,4}$ & $\parenth{4,4,1}$& $\parenth{4,4,2}$ & $\parenth{4,4,3}$& $\parenth{4,5,2}$&$\parenth{4,5,3}$&$\parenth{5,1,4}$&$\parenth{5,2,4}$&$\parenth{5,4,3}$&$\parenth{5,4,2}$&$\parenth{5,5,1}$\\[5pt]
     \hline
     $-\frac{1}{48}$& $-\frac{3}{40}$& $\frac{1}{48}$& $-\frac{1}{72}$
    & $-\frac{1}{36}$& $\frac{1}{48}$& $\frac{1}{30}$  & $-\frac{1}{72}$& $-\frac{1}{36}$& $-\frac{1}{24}$ & $\frac{1}{48}$& $-\frac{1}{24}$& $\frac{1}{48}$& $-\frac{1}{24}$& $\frac{1}{48}$& $-\frac{3}{40}$\\[5pt]
    \hline
\end{tabular}
\caption{$N=3$,$n=5$, $\lambda^{(1)}=\lambda^{(2)}=(3,2)$, $ \lambda^{(3)}=(4,1)$}
\label{tab:n5-2}
\end{table}

\begin{table}[H]
    \centering
    \begin{tabular}{|c|c|c|c|c|c|c|c|c|c|c|c|c|c|c|c|}
    \hline
         (1,1,1)& (1,2,2) & (1,2,3)& (1,3,3)& (1,4,4)&(1,4,5)&(1,5,5)&(1,6,6)&(1,7,7)&(1,7,8)&(1,8,8)&(1,9,9)&(2,2,2)&(2,2,3)&(2,3,3)&(2,4,4)\\
         \hline
         $\frac{3}{296}$& $-\frac{1}{888}$& $-\frac{25}{3996}$ & $-\frac{8}{999}$&$-\frac{1}{888}$& $-\frac{25}{3996}$&$-\frac{8}{999}$& $\frac{2}{111}$ & $-\frac{1}{888}$& $-\frac{25}{3996}$&$-\frac{8}{999}$&$\frac{2}{111}$&$\frac{5}{666}$&$-\frac{5}{2997}$&$-\frac{40}{2997}$&$-\frac{5}{2664}$\\[5pt]
         \hline
    \end{tabular}
    \begin{tabular}{|c|c|c|c|c|c|c|c|c|c|c|c|c|c|c|c|}
    \hline
         (2,4,5)& (2,4,6)&(2,5,5)& (2,5,6)&(2,7,7)&(2,7,8)&(2,7,9)&(2,8,8)&(2,8,9)&(3,4,4)&(3,4,5)&(3,4,6)&(3,7,7)&(3,7,8)&(3,7,9)&(4,4,4)\\
         \hline
         $\frac{5}{11988}$& $-\frac{5}{999}$ & $\frac{10}{2997}$& $\frac{10}{999}$&$-\frac{5}{2664}$& $\frac{5}{11988}$&$-\frac{5}{999}$& $\frac{10}{2997}$&$\frac{10}{999}$&$\frac{5}{11988}$&$\frac{10}{2997}$& $\frac{10}{999}$&$\frac{5}{11988}$&$\frac{10}{2997}$& $\frac{10}{999}$&$\frac{5}{1332}$\\[5pt]
         \hline
    \end{tabular}
   
    \begin{tabular}{|c|c|c|c|c|c|c|c|c|c|c|c|c|c|}
    \hline
         (4,4,5)& (4,4,6)& (4,5,5)& (4,5,6)& (4,7,7)&(4,7,8)&(4,7,9)&(4,8,8)&(4,8,9)&(5,7,7)&(5,7,8)&(5,7,9)&(6,7,7)&(6,7,8)\\
         \hline
         $-\frac{5}{5994}$& $-\frac{5}{1998}$&$-\frac{20}{2997}$& $\frac{5}{999}$&$-\frac{5}{1332}$& $\frac{5}{5994}$&$\frac{5}{1998}$& $\frac{20}{2997}$&$-\frac{5}{999}$&$\frac{5}{5994}$&$\frac{20}{2997}$&$-\frac{5}{999}$&$\frac{5}{1998}$&$-\frac{5}{999}$\\[5pt]
         \hline
    \end{tabular}
    \caption{$N=3$, $n=6$, $\lambda^{(1)}=\lambda^{(2)}=\lambda^{(3)}$=(4,2). The remaining basis states are generated through permutations of the labels.}
    \label{tab:n6-1}
\end{table}
\begin{table}[H]
    \centering
      \begin{tabular}{|c|c|c|c|c|}
         \hline
         (1,1,1,1)& (1,1,2,2)& (1,1,3,3)&(1,2,2,2)&(1,2,3,3)\\
         \hline
         $-\frac{1}{45}$& $\frac{1}{45}$ &$\frac{1}{45}$&$\frac{2}{45}$&$-\frac{2}{45}$\\[5pt] 
         \hline
    \end{tabular}
    \caption{$N=4$, $n=4$ $\lambda^{(1)}=\lambda^{(2)}=\lambda^{(3)}=\lambda^{(4)}=(3,1)$. The remaining basis states are generated through permutations of the labels.}
    \label{tab:N4-1}
\end{table}


%

\end{document}